\documentclass{article}

\usepackage{amssymb}
\usepackage{amsthm}
\usepackage{amsmath}

\usepackage{graphicx}

\newtheorem{theorem}{Theorem}[section]

\newtheorem{remark}[theorem]{Remark}

\newcommand{\D}{{\mathrm{d}}}

\newcommand{\wW}{{\widetilde W}}

\begin{document}
\title{Parabolic regularization of the gradient catastrophes for the Burgers-Hopf equation and Jordan chain}
\author{B. G. Konopelchenko$^1$ and G. Ortenzi$^2$\\
\small{1 Dipartimento di Matematica e Fisica 'Ennio De Giorgi'- Universit\`a del Salento and } \\
\small{\quad  INFN Sezione di Lecce, 73100 Lecce Italy } \\
\small{2 Dipartimento di Matematica e Applicazioni-Universit\`a degli Studi di Milano-Bicocca, 20125 Milano Italy} }
\date{}
\maketitle
\abstract{Non-standard parabolic regularization of gradient catastrophes for the Burgers-Hopf equation is proposed. 
It is based on the analysis of all (generic and higher order) gradient catastrophes and their step by step
regularization by embedding the Burgers-Hopf equation into multi-component parabolic systems of quasilinear
PDEs with the most degenerate Jordan block. 
Probabilistic realization of such procedure is presented. 
The complete
regularization of the Burgers-Hopf equation is achieved by embedding it into the infinite parabolic Jordan chain.
It is shown that the Burgers equation is a particular reduction of the Jordan chain. 
Gradient catastrophes for the parabolic Jordan systems are also studied.}
\section{Introduction}
The problem of gradient catastrophe (unbounded growth of derivatives) and mechanisms of its regularization have been
addressed in a number of papers in various contexts. Though the gradient catastrophe (GC) shows up in quite
different ways  in different branches of physics and mathematics, its essence is encoded in the Burgers-Hopf (BH)
(or Riemann) equation 
\begin{equation}
u_t=uu_x
 \label{BHeq}
\end{equation}
where $u=u(x,t)$ denotes a function of two independent  variables $(x,t)$
and the subscripts stand for partial derivatives. The BH equation is the most simplified one-dimensional 
version of the Navier-Stokes equation in hydrodynamics and the best studied prototype of nonlinear PDEs exhibiting
the GC (see e.g. \cite{Whi,LL6,RYa}). 
The hodograph equation $x + u t +f(u)=0$
for the BH equation provides us a simple and almost explicit formulae describing  the behavior of $u$ near
to the point of the first GC where $u$ is finite but its derivatives $u_x$ and $u_t$ are unbounded \cite{Whi,LL6,RYa}. \par
A standard method of regularization of this GC is to add higher order derivatives of $u$ in the r.h.s. 
of equation (\ref{BHeq}) \cite{Whi,LL6}. Adding $u_{xx}$, one gets the Burgers equation which mimics the effects of dissipation. 
With the $u_{xxx}$  term in the r.h.s. of (\ref{BHeq}) one obtains the Korteweg-de Vries (KdV) 
equation which describes the dispersive regularization of the BH equation. Various extensions of such types of 
regularizations have been discussed, for instance, in \cite{KS,Dubgen,DE}. \par
In the present paper we propose a method of regularization consisting in step by step regularization
of all (generic and higher order) GCs of the BH equation by introducing  new degrees of freedom. At each step
these new variables obey the parabolic system of quasi-linear partial differential equations of the Jordan block
type. The complete regularization is achieved by embedding the BH equation into the infinite Jordan chain
\begin{equation}
{u_k}_t=u_1{u_k}_x+{u_{k+1}}_x\, ,\qquad k=1,2,3,\dots\, .
 \label{Jch-intro}
\end{equation} \par
The basic point of our approach is that the complete study of the GC phenomenon for the BH equation 
has to include not only analysis of the so-called generic GC but also an investigation of the higher GCs 
with more singular behavior  of $u_x$ and $u_t$. The GC for the BH equation corresponds to the critical
point of the function $\phi(u)=x+ut+f(u)$ where $f$ is the function inverse to the initial datum $u(x,0)$.
According to the general approach formulated by Poincar\'e in \cite{Poi} (see also \cite{Arn1}) one has to study not only a single 
situation  (even generic one) but the whole family of close situations in order to get complete and deep
understanding of certain phenomenon. In our case it means that the analysis of the critical points of the 
infinite family of functions $\phi(u)$ corresponding to all possible initial data is required. The presence of
unremovable degenerate critical points of different orders is the characteristic feature of families of 
functions. This is a typical situation in the theory of singularities (or theory of catastrophes) 
(see e.g. \cite{Arn1,Tho,PS,AGV,Arn2}): 
``So, the non-generic degenerations becomes unremovable if one considers not an individual object but a family''
(\cite{AGV}, \S 8) and ``Thus, in the investigation of critical points of functions depending on parameters, we have to 
consider degenerate critical points as well as non-degenerate. A larger number of parameters
a more complicated  critical points that can occur'' \cite{Arn2}. \par

For the BH equation (\ref{BHeq}) the GC of
order $k$ corresponds to a critical point of order $k+1$ of the function $W= xu+ut^2/2+\wW(u)$,  $\D {\wW}/\D u=f(u)$
which satisfies ${\partial^l W}/{\partial u^l}=0$ for $l=0,1,\dots,k+1$ 
and is associated with the $A_{k+1}$ singularity. The derivative $u_x$ behaves like $u_x\sim (x-x_0)^{-k/(k+1)}$
near to the critical point $x_0$.
Such type of singularities, called secondary ones in the singularity theory \cite{AGV}, have important
implications in physics.
Zeldovich's theory of large-scale structure of the Universe \cite{Zel}  provides us with a notable
evidence of their relevancy. \par 
The second point is that the regularization of the GC at each step is performed by adding one new dependent
variable in a way that in a such larger space the degree of degeneration of critical point is reduced
by one. Moreover in this procedure one requires that 
1) At each step the system is parabolic in order to respect the priority role of dissipation
 and diffusion phenomena in regularization of GCs; 
 2)  The extended systems should be minimal in a sense that they should have only one characteristic
 speed as BH equation;
3)  The extended systems should include the transport equation for the BH equation
 that is the equation, for example, for the entropy $s$ in fluid-dynamics;
4) All regularizing systems should be equations governing the dynamics of critical points of
 certain functions. \par

Within this approach, the first step in the regularization of the BH equation is
\begin{equation}
 \left\{ \begin{array}{c}
          u_t=uu_x \\ s_t=us_x
         \end{array}
\right.  \longrightarrow  
 \left\{ \begin{array}{c}
          {u_1}_t={u_1}{u_1}_x+{u_2}_x \\ {u_2}_t={u_1}{u_2}_x\, .
         \end{array}
\right.
\label{s1-reg-intro}
\end{equation}
This extension regularizes the first GC of the BH equation.
In order to regularize the $N$-th order GC one adds in total $N-1$  variables and passes 
to the parabolic $N$-component system of quasilinear PDEs with the most degenerate Jordan block.
Singular sectors and associated GCs for these $N$-component Jordan systems are studied. It is shown that the dominant
behaviour of ${u_i}_x$ at the $k$-th order GC for the $N$-component case coincides with that of the $k-1$-th order 
for $N+1$-component Jordan system. \par

In this regularization process the conditions defining the $N-1$-th gradient catastrophe of the BH equation
become the equations for the critical point of a certain function $W^{(N)}(u_1,\dots, u_N)$ of $N$ variables.
It is shown that this formal procedure has an explicit realization
as the averaging of all quantities associated with the BH equation with the one-dimensional distribution
\begin{equation}
 G^{(N)}(u_1-{{u}},u_2,,\dots,u_N)=\int_{-\infty}^{+\infty}\D \lambda \,   \exp \left(i\lambda (u_1-{{u}})
 + \sum_{k=2}^N (i \lambda)^k u_k \right)\, ,
 \label{probintro}
\end{equation}
where $u_1,u_2,,\dots,u_N$ are parameters. At $N=2$ it is the standard Gaussian distribution 
which can be interpreted also as the Maxwell distribution. At $N=3$ and $N=4$, 
 $G^{(3)}$ and $ G^{(4)}$ are expressed in term of Airy and Pearcy functions, respectively. \par
 Regularization of 
 all GCs for the BH equation within this approach requires the Jordan chain (\ref{Jch-intro}) viewed as the infinite-component
 parabolic system with a single eigenvalue $u_1(x,t)$. \par
 The importance of the Jordan chain as the regularization
 of the BH equation is confirmed also by its relation with the well-known regularizing equations. First, the 
 chain (\ref{Jch-intro}) admits the reduction 
 $u_2=\frac{u_1^2}{2} +\nu {u_1}_x$,
 $u_3=\frac{1}{3}{u_1}^3+2\nu {u_1}{u_1}_x +\nu^2 {u_1}_{xx} $ and so on,
under which it becomes the Burgers equation
\begin{equation}
 {u_1}_t=2 u_1 {u_1}_x +\nu  {u_1}_{xx}\, ,
\end{equation}
that is the most canonical parabolic regularization of the BH equation \cite{Whi}.
Under this reduction the probability density (\ref{probintro}) at $N \to \infty$ is essentially
the Fourier transform of the resolvent of the adjoint Lax operator $L^*=\nu \partial_x +u_1$
for the Burgers equation. Moreover, the chain (\ref{Jch-intro}) considered formally, forgetting its parabolic
origin, admits many other differential reductions as, for example, to the KdV equation. 
This phenomenon indicating the universality of the Jordan chain as regularizator is of interest 
and requires further study.\par
The paper is organized as follows. 
In section \ref{sec-BH} the basic facts concerning GCs for the BH equation are presented. In section 
\ref{sec-BHreg} GCs for the N-component Jordan system and associated singularities are studied.
Regularization of GCs for the Jordan systems is considered in section \ref{sec-regJ}. 
Parabolic regularizations of all GCs for the BH equation by the Jordan chain  and its relation to the 
Mart\'\i nez Alonso-Shabat universal hierarchy of hydrodynamic type  \cite{MAS} are discussed in 
section \ref{sec-MAS}. Probabilistic realization of this regularization process is presented in section 
\ref{sec-prob}. The reduction of the Jordan chain to the Burgers equation is considered in section 
\ref{sec-J-B}. The universality of the Jordan chain is briefly discussed in section \ref{sec-univ}.
The Appendix contains some explicit formulae relating the BH equation and the Jordan system with two 
components as well as the discussion of the physical model of regularization with $G^{(2)}$ being the 
Maxwell distribution for velocities for the ideal Boltzmann gas 
with the temperature $T=2u_2$.
\section{Higher gradient catastrophes for the Burgers-Hopf equation}
\label{sec-BH}
Behavior of solutions of BH equation
near generic GC (when $u(x,t)$ is finite but the derivatives $u_t$ and $u_x$ are unbounded) is given by the simple and almost explicit 
well known formulae (\cite{Whi,LL6,RYa}). On the other hand the study of non-generic or higher GCs 
did attract much less attention. \par
Here we present an analysis of generic and higher GCs for the BH equation in a form appropriate  for our purposes. The BH equation (\ref{BHeq})
can be viewed (trivially)  as the equation governing the dynamics of the critical point of the function
\begin{equation}
 W(u,x,t)=xu+t \frac{u^2}{2}+\wW(u)\, .
 \label{pothodoBH}
\end{equation}
Indeed
\begin{equation}
 x+u t +\wW_u(u)=0\, .
 \label{solhodoBH}
\end{equation}
is the hodograph equation for the BH equation where $\wW_u(u)$ is connected with the initial value of $u$ at $t=0$\, . \par
Infinite families of function $W$ corresponding to the family of all possible initial data $u(x,0)$ can be viewed
as the family of functions depending on the infinite set of parameters with $W$of the form
\begin{equation}
W(x,t,t_2,\dots)= xu + t \frac{u^2}{2} +\sum_{n=3}^\infty t_{n-1} \frac{u^n}{n!} +\wW(u) \, .
\end{equation}
So the appearance of degenerate 
critical points is very natural  \cite{Arn1}-\cite{Arn2}. \par

Generic and higher GCs are associated with the strata of singular sector 
of BH equation studied in \cite{KK}. The singular sector $S$ is the union of strata $S_k$ defined 
(in a simplified way) by the conditions
\begin{equation}
 S_k= \{(x,t,u): W_u=W_{uu}= \dots \partial_u^{k+1}W=0,\,  \partial_u^{k+2}W\neq 0 \} \, , \qquad k=1,2, \dots\, .
 \label{ssec}
\end{equation}
Calculating the differential of  (\ref{solhodoBH}) or differentiating (\ref{solhodoBH}) w.r.t. $x$ and $t$, one gets in the regular sector $S_0$ ($W_{uu}\neq 0$)
\begin{equation}
 u_x=-\frac{1}{W_{uu}}\, , \qquad u_t=-\frac{u}{W_{uu}}\, .
 \label{derGC1}
\end{equation}
In the strata $S_k$ it holds
\begin{equation}
 u_x=-\frac{(k+1)!}{\partial_u^{k+2} W}\, , \qquad u_t=-\frac{(k+1)!u}{\partial_u^{k+2} W}\, .
 \label{derGCk}
\end{equation}
The first GC ($W_{uu}=0$) happens on the submanifold of variables $(x,t,t_1,t_2,\dots)$ of codimension one, and $u_x \to \infty$ 
and $u_t \to \infty$. Let us denote the points of this submanifold with $x_0,t_0,u_0$. 
Higher GCs happen on the submanifolds of codimension $k$.\par

Let us consider the multiscaling expansion around this point. 
Subtracting from general variation its infinitesimal Galilean transformation
(see e.g. \cite{Dubgen,DE})
\begin{equation}
 x\to x' = x_0-u_0 (t-t_0),
\end{equation}
one has  
\begin{equation}
 \qquad u= u_0 +\epsilon^\alpha {\upsilon}\, , \qquad t= t_0 +\epsilon^\beta {\tau}\, , x= x_0 -u_0 \epsilon^\beta {\tau}+\epsilon^\gamma {{y}}\, ,
\label{multiexp-Gali}
\end{equation}
where $\epsilon$ is a small parameter and the constants $\alpha$, $\beta$, and $\gamma$ to be fixed. Substituting  (\ref{multiexp-Gali}) in $W(u)$,
one gets at $S_k$ 
\begin{equation}
W=W(x_0,t_0,u_0) - \frac{1}{2}\epsilon^\beta {u_0}^2 {\tau}+\epsilon^\gamma u_0 {{y}} 
+\epsilon^{\alpha+\gamma} {{y}} {\upsilon} 
+ \frac{1}{2}\epsilon^{\beta+2\alpha} {\tau} {\upsilon}^2 + \epsilon^{\alpha(k+2)} A_{k+2} {\upsilon}^{k+2} +\dots\, ,
 \label{exppot}
\end{equation}
where $A_{k+2}=\frac{1}{k+2} \frac{\partial^{k+2} W}{\partial u^{k+2}}\Big{\vert}_{u_0}$.
The balance of all three nontrivial terms in (\ref{exppot}) implies that
\begin{equation}
 \alpha+\gamma=\beta+2 \alpha=\alpha(k+2)\, ,
\end{equation}
and, hence, 
\begin{equation}
 \beta=\alpha k \, , \qquad \gamma=\alpha(k+1)\, .
 \label{balanceBH}
\end{equation}
Choosing $\alpha=1$, one has 
\begin{equation}
W=W(x_0,t_0,u_0) - \frac{1}{2}\epsilon^k {u_0}^2 {\tau}+\epsilon^{k+1} u_0 {{y}} 
+ \epsilon^{k+2} W_k^* + o(\epsilon^{k+2}) \, ,
\label{potbalBH}
\end{equation}
where 
\begin{equation}
 W_k^*={{y}} {\upsilon} +\frac{1}{2}{\tau} {\upsilon}^2 +A_{k+2} {\upsilon}^{k+2}\, . 
\label{potbalBH-star}
 \end{equation}
For other balances of two (instead of three) nontrivial terms in (\ref{exppot}), namely, 
\begin{equation}
 \alpha+\gamma= \alpha(k+2)\, , \quad  \beta+2\alpha > \alpha+\gamma\, \qquad \mathrm{or} \qquad
  \beta+2\alpha= \alpha(k+2)\, , \quad  \beta+2\alpha < \alpha+\gamma\,
\end{equation}
one gets the formula (\ref{potbalBH}) with, respectively ${\tau}=0$ or ${{y}}=0$.\par
If instead of (\ref{multiexp-Gali}) one considers the na\"\i ve expansion
\begin{equation}
 \qquad u= u_0 +\epsilon^\alpha {\upsilon}\, , \qquad t= t_0 +\epsilon^\beta {\tau}\, , \qquad x= x_0 +\epsilon^\gamma {{y}}\, ,
\label{multiexp-n}
\end{equation}
one gets
\begin{equation}
W=W(x_0,t_0,u_0) - \frac{1}{2}\epsilon^\beta {u_0}^2 {\tau}+\epsilon^\gamma u_0 {{y}} 
+\epsilon^{\alpha+\gamma} {{y}} {\upsilon} +\epsilon^{\alpha+\beta} u_0 {\tau} {\upsilon} 
+ \frac{1}{2}\epsilon^{\beta+2\alpha} {\tau} {\upsilon}^2 + \epsilon^{\alpha(k+2)} A_{k+2} {\upsilon}^{k+2} +\dots\, ,
 \label{exppot-n}
\end{equation}
for $W$ of the form (\ref{exppot-n}) one has two different cases. The first, the generic one with $u_0\neq 0$.  
In this case, since $\alpha+2 \gamma > \beta +\gamma $, the balance of the other terms implies
\begin{equation}
\gamma=\beta=\alpha(k+1)\, ,
 \label{balanceBH-n}
\end{equation}
and, at $\alpha=1$
\begin{equation}
\beta=\gamma=k+1 \, .
 \label{balanceBH-n-n}
\end{equation}
The potential $W$ becomes
\begin{equation}
 W= W(x_0,t_0,u_0) +\epsilon^{k+1} \left(  u_0 {{y}} +\frac{1}{2} (u_0)^2 {\tau} \right) + \epsilon^{k+2} W^*_k
\end{equation}
with 
\begin{equation}
 W_k^*({\xi},{\upsilon})= {\xi} {\upsilon} +A_{k+2}{\upsilon}^{k+2}
\end{equation}
and ${\xi}={{y}} + u_0{\tau}$.\par

In the particular case $u_0=0$ the balance is given by
\begin{equation}
 \alpha+\gamma=2 \alpha +\beta +\alpha(k+2)\, ,
\end{equation}
and, hence, one has for $\alpha=1$, 
\begin{equation}
W=W(x_0,t_0,u_0)  + \epsilon^{k+2} W^*_k
\label{potbalBH-n-0}
\end{equation}
with
\begin{equation}
 W_k^*({\xi},{\upsilon})= {\xi} {\upsilon} +\frac{1}{2} {\upsilon}^2 {\tau}+A_{k+2}{\upsilon}^{k+2}\, .
\end{equation}
Comparing (\ref{potbalBH}), (\ref{multiexp-Gali}) and  (\ref{potbalBH-n-0}), (\ref{multiexp-n}), 
one observes that within the na\"\i ve expansion (\ref{potbalBH-n-0}) the choice $u_0=0$ is equivalent 
to the elimination of the contribution of Galilean transformation from variation of $x$.\par
The condition for the critical point for the function $ W^*_k$ is
\begin{equation}
 \frac{\partial W^*_k}{\partial {\upsilon}}= {{y}}+{\upsilon}{\tau} +(k+2) A_{k+2}{\upsilon}^{k+1}=0\, , \quad k=1,2,3,\dots
 \label{hodocat}
\end{equation}
These equations are well-known  in the cases $k=1,2$. At $k=2$ it corresponds to the first time for the so-called
generic GC for the BH equation (see e.g. \cite{Whi,LL6,Dubgen}).
Within our approach it describes first higher ($k=2$) GC with additional condition $W_{uuuu}(u_0)<0$. \par

Equation (\ref{hodocat}) defines the behavior of ${\upsilon}$  at GCs. Taking ${\tau}=0$,  one gets at $S_k$ 
${\upsilon} \sim ({{y}})^{1/(k+1)}$, and, hence
\begin{equation}
   \frac{\partial {\upsilon}}{\partial {{y}} }\sim ({{y}})^{-k/(k+1)}\, , \quad k=1,2,\dots\, .
 \label{dUbehGCk}
\end{equation}
So at higher GCs  one has more singular behavior of the derivatives.  
The behavior (\ref{dUbehGCk}) follows also
from the simple observation that at $S_k$
\begin{equation}
 u_x\sim \frac{\delta u }{ \delta x} \sim \frac{\epsilon^\alpha {\upsilon}}{ \epsilon^\gamma {{y}}} \sim \epsilon^{\alpha-\gamma} 
 \sim \epsilon^{-k \alpha}\, .
\end{equation}
Since for pure variation of $x$ we have $\delta x \sim \epsilon^{\gamma} {{y}} \sim \epsilon^{\alpha (k+1)} $ one gets
\begin{equation}
 u_x  \sim (\delta x)^{-k/(k+1)}\, .
 \label{BHder-beha}
\end{equation}
The function $W^*_k$ (\ref{potbalBH}) and equation (\ref{hodocat}) are the same as (\ref{pothodoBH}) and (\ref{solhodoBH}) 
with the particular choice $\wW^*_k=A_{k+2} {\upsilon}^{k+2}$. For each $k$, solutions of  (\ref{hodocat}) obey to the BH equation 
${\upsilon}_{{\tau}}={\upsilon}{\upsilon}_{{{y}}}$ and provide its solutions exhibiting a GC of order $k$. \par
In a similar manner one can analyze higher GCs for each member 
\begin{equation}
 u_{t_{n-1}}=\frac{1}{n!}(u^n)_x\, , \qquad n=2,3,4,\dots\, ,
 \label{BHneq}
\end{equation}
of the BH hierarchy. The $n$-th equation of the hierarchy (\ref{BHneq}) describes 
the dynamics of the critical points $\partial W_n / \partial {\upsilon}=0$
of the function
\begin{equation}
W_n(x,t_{n-1},u)=x u+ \frac{t_{n-1}}{n!} u^n +\wW_n(u)\, .
 \label{pothodoBHn}
\end{equation}
One has
\begin{equation}
\begin{split}
 &\frac{\partial W_n}{\partial u}=x+ \frac{t_{n-1}}{(n-1)!} u^{n-1}+\frac{\partial \wW_n}{\partial u}=0 \\
 &\frac{\partial^2 W_n}{\partial u^2}= \frac{t_{n-1}}{(n-2)!} u^{n-2}+\frac{\partial^2 \wW_n}{\partial u^2} \\
 &\dots
\end{split}
\label{BHnhodo}
\end{equation}
and the analogue of the formula (\ref{derGCk}) is
\begin{equation}
 u_x=-\frac{1}{\partial_u^{2} W_n}\, , \qquad u_{t_{n-1}}=-\frac{u^{n-1}}{(n-1)!\partial_u^{2} W_n}\, .
 \label{derGCkn}
\end{equation}
Singular sectors $S_k$ and $k$-th orders GCs are again defined by the formula  (\ref{ssec}). 
Performing the expansion 
\begin{equation}
u=u_0+\epsilon^{\alpha} {\upsilon}\, , \qquad  t={t_{n-1}}_0+\epsilon^{\beta} {\tau}_{n-1}\, , 
\qquad x=x_0-\epsilon^{\beta} \frac{u_0^{n-1}}{(n-1)!} {\tau}_{n-1} +\epsilon^{\gamma} {{y}}\, ,
 \label{multiexp-Gali-tn}
\end{equation}
with the balancing condition (\ref{balanceBH}), one gets ($\alpha=1$)
\begin{equation}
 W_n={W_n}(x_0,{t_{n-1}}_0,u_0) -\frac{1}{2} \epsilon^{k} (u_0)^{n-1} t_{n-1}^* + \epsilon^{k+1} u_0 {{y}} + \epsilon^{k+2} W_{n-1}^*  \, ,
\end{equation}
where
\begin{equation}
 W_{n-1}^*={{y}} {\upsilon}+\frac{1}{2(n-1)!} u_0^{n-1} {\tau}_{n-1} {\upsilon}^{k+2}+ A_{k+2}{\upsilon}^{k+2}\, .
\end{equation}
So, generically, one has the same behavior of $u$ and its derivatives $u_x,u_t$ at the GC points for all
the equations of the BH hierarchy. For the case $k=2$ see also \cite{Dubgen}. In the particular case $u_0=0$
one instead has
\begin{equation}
 W={W}(x_0,{t}_0,u_0)+ \epsilon^{\alpha+\gamma} {{y}} {\upsilon} +\frac{1}{n!} \epsilon^{n\alpha+\beta} {\tau}_{n-1} {\upsilon}^n 
 + \epsilon^{\alpha(k+2)} A_{k+2} {\upsilon}^{\alpha(k+2)} +\dots \, .
\end{equation}
So the balance is $\alpha+\gamma=n \alpha+\beta = \alpha (k+2)$ and for $\alpha=1$ we obtain
\begin{equation}
 W={W}(x_0,{t}_0,u_0)+ \epsilon^{k+2}  \left( {{y}} {\upsilon} +\frac{1}{n!} {\tau}_{n-1} {\upsilon}^n 
 + A_{k+2} {\upsilon}^{k+2}  \right)
 +\dots \, .
\end{equation}
The function $W^*_k$ (\ref{potbalBH-star}) represent a particular family of deformations for the 
singularities of $A_{k+1}$-type (see e.g. \cite{Arn1}-\cite{Arn2}). \par 

The relation between GCs for the BH equation 
and the singularities (catastrophes) of $A_{k+1}$-type is known since the paper \cite{Zee} and has been 
discussed later many times. Very interesting and important implication of the higher GCs of the BH 
equation has been discovered by Ya. B. Zel'dovich \cite{Zel}. In his model of matter distribution 
in the universe the GCs are responsible for the formation of compact objects since the density of matter $n$
is proportional to the gradient of the particle velocity near to the point $x_0$ of GC. In one-dimensional approximation
\begin{equation}
 n(x)|_{x_0} \sim u_x(x)|_{x_0}\sim (x-x_0)^{-k/(k+1)}\, .
\end{equation}
The cases $k=1,2,3,\infty$ are those considered in \cite{Zel} (see also \cite{Arn1}). 
For the recent developments of Zel'dovich's model see e.g. \cite{SZ,HSW} and references therein.
\section{Gradient catastrophe for the $N$-component Jordan system}
\label{sec-BHreg}
The BH equation (\ref{BHeq}) can be extended in many different ways following to the choice of physical effects to be incorporated or 
the general mathematical scheme in which it should be embedded. \par
Our choice is to consider the BH equation as a simplest instance of the systems of quasi-linear PDEs of the first order (hydrodynamic-type
equation)
\begin{equation}
{u_i}_t=\sum_k A_{ik}(u) {u_k}_x\, , \qquad i=1,2,\dots, N\, .
 \label{NgenBHeq}
\end{equation}
In addition to require that the characteristic speeds of the system (\ref{NgenBHeq}), i.e. eigenvalues of the matrix $A$, are all coincident.
With such a choice the system (\ref{NgenBHeq}) with a single characteristic speed is the pure parabolic system of PDEs closest to the BH equation
(\ref{BHeq}). \par
In the $2$-component case the simplest  example of such system is given by 
\begin{equation}
{u_1}_t=u_1 {u_1}_x+{u_2}_x\, , \qquad {u_2}_t=u_1 {u_2}_x\, .
 \label{2genBHeq}
\end{equation}
It is noted that the second equation (\ref{2genBHeq}) has a simple physical meaning. 
Namely, it is the transport equation for the BH
equation (\ref{BHeq}) \cite{Whi,LL6}. \par
The systems (\ref{NgenBHeq}) with the matrix $A(u)$ being the most degenerate Jordan block of the order $N$, i.e. the systems 
\begin{equation}
 \left(
 \begin{array}{c}
  u_1 \\ \vdots  \\ \vdots  \\ u_N
 \end{array}
\right)_t=
 \left(
 \begin{array}{ccccc}
  u_1 & 1 &0 & \dots& 0 \\
  0 & u_1 &1 & \dots& 0 \\
  \dots & \dots &\dots & \dots& \dots \\
  0 & 0 &\dots & u_1& 1 \\
  0 & 0 &\dots & 0& u_1 \\
 \end{array}
\right)
 \left(
 \begin{array}{c}
  u_1 \\  \vdots \\ \vdots \\ u_N  
 \end{array}
\right)_x
\label{n-Jordan}
\end{equation}
have been introduced in \cite{KK-J}. 
The $N$-component system (\ref{n-Jordan}) has a common property with the BH equation: 
namely, it has only one family of characteristics 
 \begin{equation}
  \frac{\D x}{\D t}=-u_1(x,t).
  \label{char-n-J}
 \end{equation}
Consequently it can be represented in the form
\begin{equation}
\begin{split}
\mathcal{D}_1 u_i &= \mathcal{D}_{2} u_{i+1}, \qquad i=1,2,\dots, N-1, \\
\mathcal{D}_1 u_N &=0  \, ,
\end{split}
\label{n-J-tri}
\end{equation}
with the vector fields $\mathcal{D}_1 = \partial_t -u_1\partial_x$ (derivative along the characteristic)
and $\mathcal{D}_2 = \partial_x$ which obey the relation $[\mathcal{D}_1,\mathcal{D}_2]={u_1}_x \mathcal{D}_2$.

The system (\ref{n-Jordan}) describes the dynamics of the critical points of the function \cite{KK-J}
\begin{equation}
 W^{(N)}=x u_1 +t \left( \frac{{u_1}^2}{2} +u_2\right) +\wW^{(N)}({u_1,u_2,\dots, u_N})
 \label{WN}
\end{equation}
which obey the PDEs
\begin{equation}
\frac{\partial W^{(N)}}{\partial u_k}=\frac{\partial^k W^{(N)}}{\partial u_1^k}\, , \qquad k=1,2,\dots,N\, .
 \label{WN-eqn}
\end{equation}
The hodograph equations are
\begin{equation}
 \frac{\partial W^{(N)}}{\partial u_k}=0\, , \qquad k=1,2,\dots,N\, ,
\end{equation}
i.e.
\begin{equation}
\begin{split}
 \frac{\partial W^{(N)}}{\partial u_1}&=x+ tu_1 + \frac{\partial \wW^{(N)}}{\partial u_1}=0\, , \\ 
 \frac{\partial W^{(N)}}{\partial u_2}&= t+ \frac{\partial^2 \wW^{(N)}}{\partial {u_1}^2}=0\, , \\ 
 \frac{\partial W^{(N)}}{\partial u_k}&= \frac{\partial^k \wW^{(N)}}{\partial {u_1}^k} \, , \qquad k=3,4,\dots,N\, .
\end{split}
 \end{equation}
Let us denote $ {\partial W^{(N)}}/{\partial u_k}=  W^{(N)}_k$.
One has the relation (see \cite{KK-J} sec. $6$, Lemma $2$)
\begin{equation}
 \left(
 \begin{array}{ccccc}
  W^{(N)}_{N+1} & W^{(N)}_{N+2} &\dots & \dots& W^{(N)}_{2N} \\
  0 & W^{(N)}_{N+1} & W^{(N)}_{N+2} & \dots& W^{(N)}_{2N-1} \\
  \dots & \dots &\dots & \dots& \dots \\
  0 & 0 &\dots & W^{(N)}_{N+1} & W^{(N)}_{N+2}  \\
  0 & 0 &\dots & 0& W^{(N)}_{N+1} \\
 \end{array}
\right)
 \left(
 \begin{array}{c}
  u_1 \\ u_2 \\ \vdots \\ u_N  
 \end{array}
\right)=
 \left(
 \begin{array}{c}
  p_{k-N+1} \\ p_{k-N+2} \\ \vdots  \\ p_k
 \end{array}
\right) \, , \qquad k=0,1,2,\dots\, ,
\label{KK-formula}
\end{equation}
where $t_0=x$, $t_1=t$, and $p_l$ are the standard elementary Schur polynomials defined, for nonnegative indices, by the relation
\begin{equation}
 \exp \left( \sum_{i\geq 1}^N u_i z^i \right)=\sum_{i \geq 0}^\infty p_i (u) z^i
 \label{Sch-gen}
\end{equation}
and $p_i=0$ if $i<0$. 
So
\begin{equation}
\begin{split}
 &\frac{\partial u_N}{\partial t_k} = -\frac{p_k}{W^{(N)}_{N+1}} \, , \\
 &\frac{\partial u_{N-1}}{\partial t_k} = p_k \frac{W^{(N)_{N+2}}}{(W^{(N)}_{N+1})^2}-\frac{p_{k-1}}{W^{(N)}_{N+1}}\, ,\\
 &\dots \\
 &\frac{\partial u_{l}}{\partial t_k} = C_{N-l+1} \frac{1}{(W^{(N)}_{N+1})^{N-l+1}}+C_{N-l} \frac{1}{(W^{(N)}_{N+1})^{N-l}}+
\dots C_{1} \frac{1}{(W^{(N)}_{N+1})^{N+1}}\, , \\
& \qquad l=1,\dots,N\, ,\qquad k=0,1,2,\dots\, ,
\end{split}
\label{n-Jor-k-p}
\end{equation}
with some suitable coefficient $C_m$ depending  on $u$. \par

The GC of order $k$ is defined by
\begin{equation}
 \partial_{u_1}^s W^{(N)}=0\, , \qquad s=1,2,\dots, N+k\, , \qquad \partial_{u_1}^{N+k+1} W^{(N)} \neq 0\, .
\end{equation}
The expansion around the point $x_0$, $t_0$, and ${u_i}_0$ for $i=1,2,\dots, N$ 
(with subtraction of the Galilean transformation) is given by
\begin{equation}
u_i={u_i}_0+\epsilon^{\alpha_i} \upsilon_i \, ,\quad i=1,2,\dots,N \, ,\qquad t=t_0+\epsilon^\beta {\tau}\, ,\qquad 
x=x_0-{u_1}_0 \epsilon^{\beta} {\tau}+\epsilon^\gamma {{y}}\, .
 \label{locexp}
\end{equation}
One gets
\begin{equation}
\begin{split}
 W^{(N)}=&W^{(N)}_0 +\epsilon^\gamma {u_1}_0 {{y}} +\epsilon^{\alpha_1+\gamma} {\upsilon_1} {{y}}  
 +\frac{1}{2} \epsilon^{\beta+2\alpha_1}  ({\upsilon_1})^2 {\tau}+\frac{1}{2} \epsilon^{\beta+2\alpha_2}  ({\upsilon_2})^2 {\tau}
-\frac{1}{2} \epsilon^\beta ({u_1}_0)^2 {\tau}  +\epsilon^\beta {u_0}_2 {\tau} 
\\
  & + \sum_{L} \frac{A_{N+k+1}}{l_1! \dots l_N!} \epsilon^{\alpha_1 l_1+\alpha_2 l_2+\dots+\alpha_N l_N} 
  ({\upsilon_1})^{l_1} ({\upsilon_2})^{l_2} \dots ({u_N}^*)^{l_N} 
\end{split}
  \end{equation}
where $L=\{l_k : \sum_{k=1}^N k l_k= N+k+1 \}$ and $A_l= \partial_{u_1}^k W^{(N)} |_{u_k={u_k}_0}$. 
Complete balance is achieved if
\begin{equation}
 \alpha_1+\gamma=\beta+2\alpha_1 =\beta+\alpha_2=\alpha_1(N+k+1) \, .
\end{equation}
At $\alpha_1=1$ we have
\begin{equation}
 \alpha_k=k  \quad s=1,2,\dots,N\, , \beta= N+k-1\, , 	\qquad \gamma=N+k\, .
\label{balgen-w-n-J}
 \end{equation}
and
\begin{equation}
 W^{(N)}= W^{(N)}_0+\epsilon^{N+k}{u_1}_0 {{y}} +\epsilon^{N+k-1}\left( -\frac{1}{2}({u_1}_0)^2+{u_2}_0\right) {\tau}
+\epsilon^{N+k+1} {W^{(N)}_k}^* {\upsilon} +o(\epsilon^{N+k+1})\, ,
\label{pot-n-J-k}
 \end{equation}
where 
\begin{equation}
{W^{(N)}_k}^*= {{y}} P_1({\upsilon})+{\tau} P_2({\upsilon}) +A_{N+k+1} P_{N+k+1}({\upsilon})\, . 
\label{pot-n-J-k-star}
\end{equation}
At the critical point of $W^*$ one has   
\begin{equation}
\begin{split}
 &\frac{\partial {W^{(N)}}^{*}}{\partial {\upsilon_1}}= {{y}}+{\tau} {\upsilon_1} + A_{N+k+1} P_{N+k}({\upsilon})=0\, , \\
  &\frac{\partial {W^{(N)}}^{*}}{\partial {\upsilon_1}}= {\tau} + A_{N+k+1} P_{N+k-1}({\upsilon})=0\ , , \\
 &\frac{\partial {W^{(N)}}^{*}}{\partial {\upsilon_l}}= A_{N+k+1} P_{N+k-l+1}({\upsilon})=0\, , \qquad l=3,\dots,N \, , 
\end{split}
\label{hodo-n-J-k}
\end{equation}
In particular for $N=2$
\begin{equation}
 {W^{(2)}}^*={\upsilon_1} {{y}} +\epsilon^{2}\left( \frac{1}{2}({\upsilon_1})^2+{\upsilon}_2\right) {\tau}
+\epsilon^{k+3} P_{k+3} ({\upsilon}_1,{\upsilon}_2) 
\label{pot-2-J-k} 
\end{equation}
and hodograph equations are
\begin{equation}
\begin{split}
 & {{y}} +{\upsilon_1} {\tau}+ A_{k+3} P_{k+2}({\upsilon_1},{\upsilon_2})=0\, , \\
 & {\tau} +A_{k+3} P_{k+1}({\upsilon_1},{\upsilon_2})=0\, , \qquad k=1,2,3,\dots \, .
\end{split}
\label{hodo-2-J-k}
\end{equation}
The function $(W^{(N)})^*$ and equations (\ref{hodo-n-J-k}) coincide with (\ref{WN}) and 
(\ref{WN-eqn})
with the choice $\wW^{(N)}=A_{N+k+1} P_{N+k+1}({\upsilon})$. For each $k$  solutions of equations (\ref{hodo-n-J-k})
obey the Jordan system (\ref{n-Jordan}) with the substitution $x \to {{y}}$, $t \to {\tau}$ and $u_l \to {\upsilon_l}$
and provide us with the solutions exhibit the $k$-th order GC.\par

The equation (\ref{hodo-2-J-k}) define the mapping $({\upsilon_1},{\upsilon_2}) \to ({{y}},{\tau})$ given by
\begin{equation}
 \left\{
\begin{array}{l}
 {{y}} = A_{k+3} ({\upsilon_1} P_{k+1}({\upsilon_1},{\upsilon_2})-P_{k+2}({\upsilon_1},{\upsilon_2}))\, ,\\ \\
 {\tau} = - A_{k+3} P_{k+1}({\upsilon_1},{\upsilon_2})\, .
\end{array}
 \right.
\end{equation}
At $k=1$ one has the mapping
\begin{equation}
 \left\{
\begin{array}{l}
 \frac{{{y}}}{A_4} = \frac{1}{3}({\upsilon_1})^3\, ,\\ \\
 \frac{{\tau}}{A_4} = - \frac{1}{2} ({\upsilon_1})^2 - {\upsilon_2}\, ,
\end{array}
 \right.
\end{equation}
while at $k=2$ it is
\begin{equation}
 \left\{
\begin{array}{l}
 \frac{{{y}}}{A_5} = \frac{1}{8}({\upsilon_1})^4 +\frac{1}{2} ({\upsilon_1})^2 {\upsilon_2} -\frac{1}{2} ({\upsilon_2})^2 \, ,\\ \\
 \frac{{\tau}}{A_5} = - \frac{1}{6} ({\upsilon_1})^3 - {\upsilon_1}{\upsilon_2}\, .
\end{array}
 \right.
\end{equation}
One can rewrite equations (\ref{hodo-2-J-k}) as a single equation for ${\upsilon_1}$. At $k=2$ one gets
\begin{equation}
 {{y}}-\frac{A_4}{3} ({\upsilon_1})^3=0\, ,
\end{equation}
and at $k=2$ one has
\begin{equation}
 {{{y}}} -\frac{A_5}{36} {{\upsilon_1}}^4 +\frac{({\tau})^2}{2 {\upsilon_1} A_5}+\frac{2}{3}{\upsilon_1}{\tau} =0  
\end{equation}
The formula (\ref{hodo-n-J-k}) imply that, near to the GC point
$x_0, t_0 ,u_0$  one has 
\begin{equation}
 \frac{\partial {\upsilon_l}}{\partial {{y}}} \sim ({{y}})^{-(N+k-l)/(N+k)}\, \quad l=1,2,\dots,N\,  .
 \label{sing-ul}
\end{equation}
This formula gives us the dominant (most singular) behavior of ${u_l}_x$ at the point of $k$-th order GC. Dominant and non-dominant terms can be 
obtained from the formulae (\ref{KK-formula}) and (\ref{n-Jor-k-p}). Since at $k$-th order GC $W_{N+m} \sim \epsilon^{k-(m-1)}$ for
$m=1,2,\dots, N$ and as $\delta x \to 0$, with $\alpha_1=1$, we have $\delta x \sim \epsilon^{N+k}$  one gets from (\ref{n-Jor-k-p})
\begin{equation}
 \begin{split}
  &{u_N}_x\sim \epsilon^{-k} \sim (\delta x)^{-k/(N+k)}\, , \\
  &{u_{N-1}}_x \sim d_1 \epsilon^{-k}+\sim d_2 \epsilon^{-(k+1)} \sim \tilde{d}_1 (\delta x)^{-k/(N+k)}+\tilde{d}_2 (\delta x)^{-(k+1)/(N+k)}\, , \\
  &{u_{l}}_x \sim  \sum_{i=1}^{N-l} c_i \epsilon^{-(k+i-1)} \sim \sum_{i=1}^{N-l} \tilde{c}_i (\delta x)^{-(k+i-1)/(N+k)} \, , \qquad l=1,2,\dots,N\, ,
  \end{split}
\label{tri-bal}
  \end{equation}
where $d,\tilde{d},c,\tilde{c}$ are certain constants. \par
The presence of non-dominant (less singular terms) in the 
formulae describing the behavior of derivatives ${u_l}_t$ at the point of GCs allows to equilibrate the singular terms of different orders in 
equations (\ref{n-Jordan}). It is a novel and important feature of multi-component system (\ref{n-Jordan}) in comparison with the BH equation 
({\ref{BHeq}}). \par 
In order to recover these non-dominant terms within the multiscale expansion one has to modify the formulae (\ref{locexp}) and (\ref{balgen-w-n-J}).
Namely, in the generic case ${u_1}_0\neq 0$, one should consider the expansion near the GC point of order $k$ in the form 
\begin{equation}
x=x_0 + \epsilon^{N+k}{{y}}\, , \qquad t=t_0 + \epsilon^{N+k}{\tau}\, , \qquad {u_l}={u_l}_0 + \sum_{m=l}^N \epsilon^{m} {\upsilon_{lm}}\, ,
 \label{loctribal}
\end{equation}
for $l=1,2,\dots, N$. Using (\ref{loctribal}), one immediately gets the formulae (\ref{tri-bal}) for ${u_l}_x\sim \frac{\delta u_l}{ \delta x}$.
Then, straightforward calculations show that the non-leading terms ${\upsilon_{lm}}$, $m=l+1, \dots, N$ in the formulae (\ref{loctribal}) for $u_l$
give contributions only for terms of order $\epsilon^{N+k+2}$ and higher in the expansion of $W(x,t,u)$ near to the point
$x_0,t_0,u_0$, leaving unchanged the dominant term of order $\epsilon^{N+k+1}$, namely 
$\epsilon^{N+k+1} {W^{(N)}_k}^*({\upsilon}_{11}, \dots, {\upsilon}_{NN})$. \par
For example in the simplest case $N=2$ and $k=1$ one has
\begin{equation}
x=x_0-u_{10} \epsilon^2 \tau + \epsilon^3{{y}}\, , \qquad t=t_0 + \epsilon^2 {\tau}\, , \qquad {u_1}={u_1}_0 + \epsilon \upsilon_{11}+ \epsilon^2 \upsilon_{12}\, ,
\qquad {u_2}={u_2}_0 + \epsilon^2 \upsilon_{22} \, ,
 \label{loctribal-2f}
\end{equation}
and 
\begin{equation}
 \begin{split}
  W=&W(x_0,t_0,{u_1}_0,{u_2}_0)+\epsilon^3 \left( {u_1}_0 {{y}} +\left(\frac{1}{2}{u_1}_0^2+{u_2}_0\right){\tau} \right)
  +\epsilon^4 \left( \xi \upsilon_{11} +A_4 P_4 (\upsilon_{11},\upsilon_{22})\right)\\
  &+\epsilon^5 \left(\left( \xi  +A_4 P_4 (\upsilon_{11},\upsilon_{22})  \right) \upsilon_{12}+ A_5 P_5 (\upsilon_{11},\upsilon_{22})  \right) +\dots \, .
 \end{split}
\end{equation}
One has similar situation in the particular case $u_{10}=0$. \par
Finally we note that due to the formulae (\ref{sing-ul}) and (\ref{tri-bal}) the dominant
(most singular) behavior of the derivative ${u_l}_x$ near the point of GC is the same for given $N+k$. In particular, the dominant singularity
of ${u_l}_x$ at $k$-th order GC for $N$-component case coincides with that of $k-1$-th order GC for $N+1$-component case. At
$k$-th order GC for two-component case the derivative ${u_1}_x$ dominantly behaves as that of ${u_1}_x$ for the first order GC at $k+1$-component
system.
\section{Regularization of gradient catastrophes for the Jordan systems}
\label{sec-regJ}
The observation made at the end of previous section on the disappearance of the first and subsequent GCs after the increasing the number
of dependent variables $u_l$ suggests a way to regularize GCs for Jordan systems. \par 

Let us start with the two component system
\begin{equation}
 {u_1}_t=u_1 {u_1}_x +{u_2}_x\, , \qquad {u_2}_t=u_1 {u_2}_x\, . 
 \label{para-toy}
\end{equation}
It describes the dynamics of the critical points of the function
\begin{equation}
 W^{(2)}(x,t,u_1,u_2)=xu+t\left( \frac{u_1^2}{2}+u_2\right)+\wW^{(2)}(u_1,u_2)
 \label{pot2}
\end{equation}
where $W^{(2)}(x,t,u_1,u_2)$ (and consequently $\wW^{(2)}(u_1,u_2)$) satisfies the heat equation
\begin{equation}
 W^{(2)}_{u_2}=W^{(2)}_{u_1 u_1}\, .
 \label{heat}
\end{equation}
The first GC happens when 
\begin{equation}
 W^{(2)}_{u_1 u_2}=W^{(2)}_{u_1 u_1 u_1}=0\, , 
 \label{1GC-toy}
\end{equation}
and hence the second differential 
\begin{equation}
 \D^2 W^{(2)}= \sum_{i,k=1}^2  \frac{\partial^2 W^{(2)}}{\partial u_1 \partial u_k}\D u_i \D u_k
\end{equation}
is degenerate, i.e.
\begin{equation}
 \left\vert
 \begin{array}{cc}
  W^{(2)}_{u_1 u_1} &  W^{(2)}_{u_1 u_2} \\  W^{(2)}_{u_2 u_1} &  W^{(2)}_{u_2 u_2} 
 \end{array}
 \right\vert = 0\, . 
\end{equation}
A way to avoid this degeneracy is to expand our two dimensional space $(u_1, u_2)$ to the three dimensional one with local coordinates
$(u_1, u_2, u_3)$ and to consider there a function $W^{(3)}(u_1,u_2,u_3)$ which, in addition to equation (\ref{heat}), i.e.  
\begin{equation}
 W^{(3)}_{u_2}=W^{(3)}_{u_1 u_1 }\, ,
 \label{2heat3}
\end{equation}
obeys also to the 
equation
\begin{equation}
 W^{(3)}_{u_3}=W^{(3)}_{u_1 u_1 u_1}\, .
 \label{3heat3}
\end{equation}
In this $3$-dimensional space the condition (\ref{1GC-toy}) of the first GC at the case $N=2$ becomes part of the equations
\begin{equation}
W^{(3)}_{u_1} =W^{(3)}_{u_2} =W^{(3)}_{u_3} =0\, ,
\end{equation}
defining critical point of function $W^{(3)}$. The second differential
\begin{equation}
 \D^2 W^{(3)}= \sum_{i,k=1}^3  \frac{\partial^2 W^{(3)}}{\partial u_1 \partial u_k}\D u_i \D u_k
\end{equation}
is non-degenerate since
\begin{equation}
 \left\vert
 \begin{array}{ccc}
  W^{(3)}_{u_1 u_1} &  W^{(3)}_{u_1 u_2}&  W^{(3)}_{u_1 u_3} \\  W^{(3)}_{u_2 u_1} &  W^{(3)}_{u_2 u_2}&  W^{(3)}_{u_2 u_3}
  \\  W^{(3)}_{u_3 u_1} &  W^{(3)}_{u_3 u_2}&  W^{(3)}_{u_3 u_3}
 \end{array}
 \right\vert = 
  \left\vert
 \begin{array}{ccc}
  0 & 0 & \partial_{u_1}^4 W^{(3)} \\  0 &  \partial_{u_1}^4 W^{(3)} & \partial_{u_1}^5 W^{(3)}
  \\  \partial_{u_1}^4 W^{(3)} &  \partial_{u_1}^5 W^{(3)}&  \partial_{u_1}^6 W^{(3)}
 \end{array}
 \right\vert 
 =  (\partial_{u_1}^4 W^{(3)})^3 \neq 0\, , 
\end{equation}
in the regular sector for $3$-component Jordan system defined by $\partial_{u_1}^4 W^{(3)}\neq 0$.\par
So the addition of the new variable $u_3$ and transition from the function $W^{(2)}$ to the function $W^{(3)}$ 
obeying (\ref{2heat3}) and (\ref{3heat3}) and 
consequently from the system (\ref{para-toy}) to the system
\begin{equation}
\left( \begin{array}{c}
        u_1 \\ u_2 \\u_3
       \end{array}
\right)_t=
\left( \begin{array}{ccc}
        u_1 & 1 & 0 \\ 0 & u_1 & 1 \\ 0 & 0 & u_1
       \end{array}
\right)
\left( \begin{array}{c}
        u_1 \\ u_2 \\u_3
       \end{array}
\right)_x
 \label{3-paratoy}
\end{equation}
regularizes the first GC for the system (\ref{para-toy}).\par 

However the system (\ref{3-paratoy}) has its own GCs. The first one appears when $\partial_{u_1}^4 W^{(3)}= 0$. In order to regularize it we add 
new variable $u_4$ and consider the functions $W^{(4)}(u_1,u_2,u_3,u_4)$ obeying the conditions
\begin{equation}
 W^{(4)}_{u_k}= \partial_{u_1}^4 W^{(4)}\, , \qquad k=2,3,4\, .
\end{equation}
The condition $\partial_{u_1}^4 W^{(3)}=0$ becomes the criticality condition $W^{(4)}_{u_4}=0$ and the second differential is 
nondegenerate in the regular sector  $\partial_{u_1}^5 W^{(4)}\neq0$. 
Repeating this process iteratively, one arrives at the $N$-component Jordan system 
(\ref{n-Jordan}) for which the condition for critical points is (\ref{WN-eqn}), i.e.  
\begin{equation}
 \partial_{u_1}^k W^{(N)}=0\, , \qquad k=1,2,\dots, N\, .
\end{equation}
The system (\ref{n-Jordan}) represents the regularization of the first order GCs for all the Jordan systems with a smaller number of fields
(from two to $N-1$).\par
Moreover, it provides us with the regularization of the higher order GCs for these systems. Indeed let us consider again the system (\ref{para-toy}).  
Its second order GC is verified if 
\begin{equation}
 \partial_{u_1}^3 W^{(2)}=\partial_{u_1}^4 W^{(2)}=0\, .
\end{equation}
Its regularization is achieved by expanding the two dimensional space $(u_1,u_2)$ to the $4$-dimensional space with coordinates 
$(u_1,u_2,u_3,u_4)$ and considering the critical powers of the function $W^{(4)}(u_1,u_2,u_3,u_4)$ obeying the equations
\begin{equation}
W^{(4)}_{u_k}= \partial_{u_1}^k W^{(4)}\, , \qquad k=2,3,4\, .
\end{equation}
Critical points of such functions are governed by the system (\ref{n-Jordan}) with $N=4$ and the second differential 
of the potential $W^{(4)}$ is regular when $\partial_{u_1}^5 W^{(4)}\neq 0$. In order to regularize $k$-th order GCs for the system 
(\ref{para-toy}) one introduces $k$ new variables $u_3,\dots,u_{k+2}$ and a function $W^{(k+2)}(u_1,\dots,u_{k+2})$ obeying the equations
\begin{equation}
 W^{(k+2)}_{u_l}=\partial^l_{u_1} W^{(k+2)}\, , \qquad l=2,\dots,k+2. 
 \label{req}
\end{equation}
Its critical points are governed by the system (\ref{n-Jordan}) with $N=k+2$ which provides us with 
the regularization of all GCs for system (\ref{para-toy}) of orders $1,\dots,k$. 
In a similar way one regularize higher GCs for the system (\ref{3-paratoy})  and so on. Thus,
the system (\ref{n-Jordan}) gives us  the regularization of GCs of different orders for the Jordan systems with a smaller number of 
components. \par
In order to regularize the GCs of all orders one should proceed to the formal limit $N \to \infty$. 
In this limit one has the infinite Jordan system \cite{KK-J}
\begin{equation}
{u_l}_t = u_1 {u_l}_x + {u_{l+1}}_x\, , \qquad l=1,2,3,\dots \, .
 \label{inf-Jordan}
\end{equation}
i.e. the Jordan chain (\ref{Jch-intro}). \par
We would like to emphasize that the method of regularization of singularities of different types by expanding the space of variables 
is well-known and has been widely used in various branch of mathematics and physics (e.g. blows-up). Here it is realized
for a specific system and under specific requirements. 
\section{Parabolic regularization of gradient catastrophes for the Burgers-Hopf equation by the Jordan chain}
\label{sec-MAS}
Here we turn back to the BH equation (\ref{BHeq}) and will apply to it the method of regularization of GCs by adding new degrees of freedom
(expansion of the space of dependent variables). We intend to apply this method  within a sort of minimalistic approach. Minimalistic in the following 
sense:
\begin{enumerate}
 \item In physics dissipation and viscosity are the principal phenomena which regularize GC \cite{Whi,LL6}. Both of them are of parabolic nature.
 So our first requirement is that the extensions (regularizations) of the BH equation to 
 larger systems should belong to the class of parabolic systems 
 of PDEs of the first order;
 \item The extended systems should incorporate the transport equation for the BH equation, i.e., 
 the equation $s_t=us_x$ which is one of the 
 important equations in hydrodynamics. For example, it is the equation for entropy in adiabatic processes \cite{Whi,LL6};
 \item The extended systems should be minimal in the sense that should have the same number of characteristic speeds as the BH equation, i.e.
 only one;
 \item Such extended systems can be viewed as those governing the dynamics of critical points of certain functions.
\end{enumerate}

Within this approach the first and simplest step of extension of the BH equation is the following one
\begin{equation}
 \left\{ \begin{array}{c}
          u_t=uu_x \\ s_t=us_x
         \end{array}
\right.  \longrightarrow  
 \left\{ \begin{array}{c}
          {u_1}_t={u_1}{u_1}_x+{u_2}_x \\ {u_2}_t={u_1}{u_2}_x\, .
         \end{array}
\right.
\label{s1-reg}
\end{equation}
Note that due to the invariance of the transport equation under the transformation $u_2 \to f(u_2)$ with arbitrary function $f(u_2)$,
the equation ${u_1}_t={u_1}{u_1}_x+a(u_2){u_2}_x$ with any function $a$ is equivalent to that in (\ref{s1-reg}). 
The system (\ref{s1-reg}) describes the dynamics of the critical points of the functions
\begin{equation}
 W^{(2)}(u_1,u_2)=xu_1 +t \left( \frac{1}{2} u_1^2 +u_2\right) +\wW^{(2)}(u_1,u_2)
\end{equation}
where $W_2$ obeys the parabolic equation
\begin{equation}
 {W}^{(2)}_{u_2}={W}^{(2)}_{u_1u_1}\, .
\end{equation}
The condition ${W}_{u u }=0$ of the first GC for the BH equation becomes the condition ${W}^{(2)}_{u_1u_1}={W}^{(2)}_{u_2}=0$ of the critical point 
for $W^{(2)}$ and the degenerate second order differential $ \D W = W_{uu} (\D u)^2$
%
becomes nondegenerate 
\begin{equation}
 \D W^{(2)} = \sum_{i,j=1}^2W_{u_iu_j}^{(2)} \D u_i \D u_j\, .
\end{equation}\par
Thus, the first gradient catastrophe of the BH equation is regularized by the passage given in (\ref{s1-reg}). \par

However, this extension does not regularize the second GC of the BH equation. Comparing (\ref{BHder-beha}) at $k=2$ and 
(\ref{sing-ul}),(\ref{tri-bal}) with $N=2,l=1$, we observe that ${u_1}_x$ at the second GC for the BH equation has the same singular 
behavior $(\delta x)^{-2/3}$ as ${u_1}_x$ at the first GC for the two-component Jordan system. So, in order to regularize the second 
GC for the BH equation one should add two new dependent variables to the BH equation or one new variable to the two component system 
(\ref{s1-reg}). In both ways one ends up with the $3$-component Jordan system. \par

Continuing this process one regularizes the first $N-1$ GCs of the BH equation passing to the $N$-component Jordan system (\ref{n-Jordan}). 
In order to regularize all  GCs for the BH equation one should pass formally to the infinite Jordan system  (\ref{inf-Jordan})
\begin{equation}
 {u_l}_t = u_1 {u_l}_x + {u_{l+1}}_x\, , \qquad l=1,2,3,\dots \, .
 \label{inf-J}
\end{equation}
This Jordan chain represents the minimal parabolic regularization of the BH equation (\ref{BHeq}). \par

In fact,
the infinite Jordan system (\ref{inf-J}) has been derived in different context fifteen years 
ago. 
As the infinite chain for the moments of the finite component $\epsilon$-systems it has been obtained in \cite{Pav}. It
appeared also as a very particular example within the classification of integrable hydrodynamic chains \cite{OS}.
But for the first time the infinite system (\ref{inf-Jordan}) appeared (in a 
equivalent form) in \cite{MAS} as the first flow  
\begin{equation}
 \frac{\partial y_n}{\partial t_1}=y_1 \frac{\partial y_n}{\partial x}- y_n \frac{\partial y_1}{\partial x}+\frac{\partial y_{n+1}}{\partial x}
\, , \qquad n=1,2,3,\dots 
\label{MAS-f}
 \end{equation}
of the universal hierarchy of hydrodynamical type. Indeed, one can show that
the chains  (\ref{inf-Jordan}) and (\ref{MAS-f}) are connected by the simple 
invertible change of variables
\begin{equation}
 y_n=p_n(u_1,u_2,\dots)\, , \qquad n=1,2,3,\dots 
 \label{J-MAS-cov}
\end{equation}
where $p_n$ are the elementary Schur polynomials with infinite set of variables. The relation (\ref{J-MAS-cov}) relates both infinite hierarchies
too. \par 

The Mart\'\i nez Alonso-Shabat (MAS) hierarchy has been widely studied 
\cite{MAS,MAS2,Sha}.  For example, it was shown that the finite-component reductions
of the MAS chain obtained imposing $y_l=0 $ for $l>N$  represent itself a diagonalizable integrable hydrodynamic systems.
It should be noted that due to the relation
$ \exp \left( \sum_{n=1}^\infty u_n z^n\right) = \sum_{n=0}^\infty y_n z^n$,
the above finite-component reduction of the MAS chain is equivalent to the algebraic reduction of the infinite
Jordan chain. Viceversa, the $N$-component Jordan system is an algebraic reduction of the infinite MAS chain.
It seems that such a reduction of the MAS chain has not been studied before. 
\section{Parabolic regularizations as averaging with the generalized Airy distribution}
\label{sec-prob}
Formal regularization procedure described in the previous sections has an explicit probabilistic
realization. First one observes that the general solution of the system of PDEs (\ref{WN-eqn}) 
can be written as
\begin{equation}
 W^{(N)}(u_1,\dots,u_N)= 
 \int_{-\infty}^{+\infty} \D \lambda \, f(\lambda) \exp\left(\sum_{k=1}^N (i\lambda)^k u_k \right) \, ,  
 \label{GATF}
\end{equation}
where $f(\lambda)$ is an arbitrary function. Equations (\ref{WN-eqn}) represent a subsystem of PDEs
which define the generalized Airy function considered in \cite{GRS}. We will refer  to the function (\ref{GATF})
as generalized Airy-type functions. Since 
\begin{equation}
W^{(1)}(u_1) \equiv  W^{(N)}(u_1,0,\dots,0)= 
 \int_{-\infty}^{+\infty} \D \lambda\, f(\lambda) \exp\left( i\lambda u_1\right) \,  ,  
 \label{GATF000}
\end{equation}
one can rewrite (\ref{GATF}) as
\begin{equation}
 W^{(N)}(u_1,\dots,u_N)= 
 \int_{-\infty}^{+\infty}  \D {{u}}\, W^{(1)}({{u}}) G^{(N)}(u_1-{{u}}, u_2, \dots, u_N ) \, ,  
 \label{potN-mean}
\end{equation}
where
\begin{equation}
 G^{(N)}(u_1-{{u}},\dots,u_N)=  \frac{1}{2\pi}
 \int_{-\infty}^{+\infty} \D \lambda\, \exp\left(i \lambda (u_1-{{u}}) +\sum_{k=2}^N (i\lambda)^k u_k \right)\,  .
 \label{N-dens}
\end{equation}
For even $N$ the function $G^{(N)}(u_1-u,u_2,\dots,u_N)$ is well defined for $u_N>0$ if $N=4n+2$  and $u_N<0$ if $N=4n$ 
with $n=0,1,2,\dots$. \par  
One immediately  has 
\begin{equation}
 \int_{-\infty}^{+\infty} \D u \, G^{(N)}(u_1-{{u}}, u_2, \dots, u_N ) =1\, , \qquad N=1,2,\dots\, .
\end{equation}
So all $G^{(N)}$ represent densities of one-dimensional probability distributions depending on $N$  
parameters $u_1,u_2,\dots,u_N$. The function $W^{(N)}(u_1,\dots,u_N)$ is the mean value of $W^{(1)}(u)$
\begin{equation}
W^{(N)}(u_1,\dots,u_N)= \langle W^{(1)}(u) \rangle_N\, .
\label{partfun}
\end{equation}
In particular
\begin{equation}
\langle \frac{u^n}{n!} \rangle_N= p_n(u_1,\dots,u_N) \, .
\end{equation}
Indeed 
\begin{equation}
 \begin{split}
\langle \frac{u^n}{n!} \rangle_N=&
 \frac{1}{2\pi n!} \int_{-\infty}^{+\infty}  \D {{u}} ({{u}})^n \int_{-\infty}^{+\infty} \D \lambda
\exp\left( -i \lambda {{u}} +\sum_{k=1}^N (i \lambda)^k u_k\right) \\
=& 
 \frac{1}{2\pi n!} \int_{-\infty}^{+\infty} \D \lambda \exp\left(\sum_{k=1}^N (i \lambda)^k u_k\right) 
  i^n \frac{\partial^n}{\partial \lambda^n } \int_{-\infty}^{+\infty}  \D {{u}} 
 \exp\left( -i \lambda {{u}} \right) \\
 =& \frac{(-i)^n}{n!} \int_{-\infty}^{+\infty} \D \lambda\,  \delta(\lambda) i^n \frac{\partial^n}{\partial \lambda^n}  
 \exp\left(\sum_{k=1}^N (i \lambda)^k u_k\right) 
    =
p_n(u_1,\dots,u_N) \, .
 \end{split}
\end{equation}
where $\delta(\lambda)$ is the Dirac-delta function. Consequently for
\begin{equation}
 W^{(1)}_m (u) = \sum_{k=1}^m \frac{u^k}{k!} t_k +\wW_1(u)
\end{equation}
one has
\begin{equation}
 \langle W^{(1)}_m  \rangle_N = \sum_{k=1}^m t_k p_k(u_1\, \dots, u_N) +\wW_m (u_1,\dots,u_N)
 = W^{(N)}_m(u_1,\dots,u_N)\, ,
\end{equation}
where 
\begin{equation}
 \wW (u_1,\dots,u_N;m) \equiv \langle W^{(1)}_m  \rangle_N\, .
\end{equation}
Moreover, using as before the integration by parts, one obtains
\begin{equation}
\frac{\partial W^{(N)}_m(u_1,\dots,u_N)}{\partial u_k}
= \langle \frac{\partial W^{(1)}(m)}{\partial u^k}  \rangle_N\, , \qquad k=1,\dots,N\, ,
\label{corr}
\end{equation}
for the integrable functions $W^1(u)$ such that $W^1(u)G^N(u_1-u,u_2,..,u_N)\to 0$ as $|u|\to \infty$.
The equations (\ref{partfun}) and (\ref{corr}) show that the GC conditions of order $N-1$ (\ref{ssec}) 
for the BH equation after averaging with probability $G^{(N)}$ become the equations defining
critical points of the function $W^{(N)}$.
Dynamics of these critical points given by the $N$-component Jordan system represent
the averaging of the BH equation and of the whole BH hierarchy (\ref{BHneq}). 

This regularization of the BH equation by averaging considered here  is similar to the classical
Whitham's averaging method {\cite{Whi2,Whi}}. We note that if one, again analogously to the Whitham's 
averaging of conservation laws \cite{Whi}, performs averaging of equation $u_t =(u^2/2)_x$, i.e.
the equation (\ref{BHeq}) by the substitution $u\to \langle u \rangle_N$,
$u^2/2\to \langle u^2/2 \rangle_N$, one gets only the first equation of the Jordan system (\ref{n-Jordan}),
while the averaging of the function $W^{(1)}(u)$ given by (\ref{partfun}) provides us with the complete system (\ref{n-Jordan}).

The density $G^{(N)}$ of the probability distribution plays a central role in such parabolic
regularizations. It can be written in different ways. Using the formal relation (\ref{Sch-gen}), one gets
\begin{equation}
 G^{(N)}(u_1-{{u}}, \dots , u_N) 
 =\frac{1}{2 \pi}\int_{-\infty}^{+\infty} \D \lambda\,  \exp(-i \lambda {{u}}) 
 \sum_{k=0}^\infty (i\lambda)^k p_k
 =  \sum_{k=0}^\infty (-1)^k p_k(u_1,\dots,u_N) \delta^{(k)}({{u}}) .
\end{equation}
where $\delta^{(k)}({{u}})$ is the $k-th$ derivative of the Dirac-delta function.
So 
\begin{equation}
 W^{(N)}(u_1,\dots, u_N)=\sum_{k=0}^\infty 
  \frac{\partial^k W^{(1)}({{u}}) }{\partial ({{u}})^k} \Big{|}_{{{u}}=0} p_k(u_1,\dots,u_N)\, .
\end{equation}
For $N=1,2,3$ one has simples explicit expressions for $G^{(N)}$.
At $N=1$ one obviously has 
\begin{equation}
 G^{(1)}(u_1-{{u}})= \delta(u_1-{{u}})\, .
\end{equation}
For $N=2$ and $u_2>0$ one easily gets
\begin{equation}
 G^{(2)}(u_1-{{u}},u_2)= \frac{1}{\sqrt{4\pi u_2}} \exp\left( -\frac{(u_1-{{u}})^2}{4u_2}\right).
 \label{heatfund}
\end{equation}
So at in this case one obtain the Gau{\ss} probability distribution with $u_1$ being the maximum 
and $2u_2$ being the squared width. Thus the regularization of the first $GC$ for the BH equation
is achieved by averaging with the Gau{\ss} distribution with finite width $\sqrt{2u_2}$.\par 
In the case $N=3$ one proceeds in standard manner using the identity
\begin{equation}
 i(u_1-{{u}}) \lambda -\lambda^2 u_2-i\lambda^3 u_3=
\frac{2 {u_2}^3}{27{u_3}^2}- \frac{(u_1-{{u}}) {u_2}^2}{3{u_3}}
+i\left( \frac{u_1-{{u}}}{(3u_3)^{1/3}}- \frac{u_2}{(3u_3)^{4/3}} \right)\lambda'
 -i \frac{\lambda'^3}{3}
\end{equation}
where
\begin{equation}
 \lambda'=(3u_3)^{1/3}\lambda -\frac{iu_2}{(3u_3)^{2/3}}\, .
\end{equation}
One gets
\begin{equation}
G^{(3)}= \frac{1}{(3u_3)^{1/3}} 
\exp\left(\frac{2 {u_2}^3}{27{u_3}^2}- \frac{(u_1-{{u}}) {u_2}^2}{3{u_3}}\right)\times
\mathrm{Ai}\left( \frac{u_2}{(3u_3)^{4/3}}- \frac{u_1-{{u}}}{(3u_3)^{1/3}} \right)
\label{GAiry}
\end{equation}
where Ai is the Airy function
\begin{equation}
 \mathrm{Ai}(z)=
 \frac{1}{2\pi}\int_{-\infty}^{+\infty}  \D \lambda\, \exp\left( -i\lambda z -i\frac{\lambda^3}{3} \right) \, .
\end{equation}
For $N=4$ one gets 
 \begin{equation}
  \begin{split}
  G^{(4)}&(u_1-{{u}},u_2,u_3,u_4)= \\
  &= 
  \frac{1}{a}\exp \left( \frac{3 {u_3}^4}{256 a^{12}}+\frac{{u_2} {u_3}^2}{16 a^8}+\frac{({u_1-u}) {u_3}}{4 a^4} \right) 
\times  \Lambda \left( 
\frac{3 {u_3}^2}{8 a^6}+\frac{{u_2}}{a^2}
,\frac{{u_3}^3}{8 a^9}
+\frac{{u_2} {u_3}}{2 a^5}
+\frac{{u_1-u}}{a}
 \right)
  \end{split}
  \label{GPearcy}
 \end{equation}
where $a=(-u_4)^{1/4}$ and
 \begin{equation}
  \Lambda(x,y)= 
  \frac{1}{2\pi}\int_{-\infty}^{+\infty}  \D \lambda\, 
 \exp\left( {i \lambda} y    -{\lambda^2} x -{\lambda^4}\right) \,.
 \end{equation}
is a Pearcey function.
So the regularization of second and third order GCs for the BH equation is achieved by averaging
with the Airy and Pearcey distributions (\ref{GAiry}) and (\ref{GPearcy}). \par
The functions $G^{(N)}(u_1-u,u_2,\dots,u_N)$ decay exponentially like exp$\left( -u^{N/(N-1)}\right) u^a$ (with a suitable $a$) at one of the
infinities (depending on $N$). 
So the regularization of the $k$-th order GC for the BH equation with $u_x \sim (\delta x)^{-k/(k+1)}$ (\ref{BHder-beha}) requires the averaging
with the probability densities which decrease at one of the infinites as exp$\left( -u^{(k+1)/k}\right)$.
\section{Jordan chain and Burgers equation}
\label{sec-J-B}
The Burgers equation 
\begin{equation}
 \phi_t=2\phi \phi_x+\nu \phi_{xx}\, ,
 \label{Burgeq}
\end{equation}
is the most known and straightforward regularization for the BH equation due to the effects of viscosity and dissipation \cite{Whi,LL6}. Here we will
show that the regularization of BH equation provided by the Jordan chain (\ref{inf-Jordan}) and Burgers equation are strictly 
connected. \par

Consider the hierarchy of infinite Jordan systems \cite{KK}, the first member of which is given by (\ref{inf-Jordan}). Note that 
for the higher Jordan flows  one has
\begin{equation}
 \frac{\partial u_1}{\partial t_k}=p_k {u_1}_x +p_{k-1} {u_2}_x +\dots +{u_{k+1}}_x =\frac{\partial p_{k+1}}{\partial x}\, , k=2,3,4,\dots\, ,
\label{n-J-P}
 \end{equation}
where $p_k$ elementary Schur polynomials. The first equation (\ref{n-Jordan})
\begin{equation}
 {u_1}_t =  {u_1}  {u_1}_x + {u_2}_x\, , 
 \label{c-JB}
\end{equation}
is obviously converted into the Burgers equation (\ref{Burgeq}) under the differential constraint
\begin{equation}
 u_2=\frac{u_1^2}{2} +\nu {u_1}_x\, .
 \label{c-JB-n}
 \end{equation}
The rest of equations (\ref{n-Jordan}) defines all other $u_l$ fields in terms of $u_1$, namely,
\begin{equation}
 \begin{split}
  u_3=&\frac{1}{3}{u_1}^3+2\nu {u_1}{u_1}_x +\nu^2 {u_1}_{xx}  \\
  u_4=& \frac{1}{4} {u_1}^4 +3 \nu {u_1}^2{u_1}_x +3 \nu^2 {u_1}{u_1}_{xx}+\frac{5}{2} \nu^2 {u_1}_x^2+ \nu^3 {u_1}_{xxx}  \\
  \dots & \, .
 \end{split}
\label{u3red}
\end{equation}
So the constraints (\ref{c-JB}) and (\ref{c-JB-n}) define an admissible reduction of the Jordan chain
(\ref{n-Jordan}) into the single
Burgers equation. \par
Further, the elementary Schur polynomials under this reduction becomes the differential polynomials defined by recursion
\begin{equation}
 p_{k+1}=(\nu \partial_x+u_1)^k u_1\, , \qquad k=1,2,3,\dots\, .
\end{equation}
Hence, under the constraints (\ref{c-JB}) and (\ref{c-JB-n}) the equations (\ref{n-J-P}) become
\begin{equation}
\frac{\partial  u_1}{\partial t_k}=\left( (\nu \partial_x+u_1)^k u_1 \right)_x\, , \qquad k=1,2,3,\dots
\end{equation}
that is nothing but the well-known 
infinite Burgers hierarchy. Thus, the whole hierarchy of Jordan chains is converted into the unconstrained 
hierarchy of Burgers equations. The fact that the universal hydrodynamic type hierarchy admits the reduction
to the Burgers hierarchy has been observed for the first time in \cite{MAS}. \par
Note that the generating function (\ref{Sch-gen}) of the elementary Schur polynomial for the Jordan chain  under the reduction (\ref{c-JB-n}) becomes 
\begin{equation}
 G= \sum_{k=0}^\infty \left(z (\nu \partial_x+u_1)\right)^k  \cdot 1 
 = \left(1-z (\nu \partial_x+u_1) \right)^{-1} \cdot 1\, .
\end{equation}
In this Burgers reduction the distribution $G^{(N)}$ is given by
\begin{equation}
 G^{(\infty)}(u_1-{{u}})=\frac{1}{2\pi} 
 \int_{-\infty}^{+\infty}  \D \lambda\, e^{-i\lambda {{u}}} 
 \left( 1-i\lambda(\nu \partial_x+u_1)\right)^{-1}\cdot 1  \, , 
 \label{Grinf}
\end{equation}
and
\begin{equation}
 W^{(\infty)}(u_1)=\frac{1}{2\pi} 
 \int_{-\infty}^{+\infty}  \D \lambda\, f({\lambda}) 
 \left( 1-i\lambda(\nu \partial_x+u_1)\right)^{-1}\cdot 1  \, .
\label{potinf}
 \end{equation}
The operator $(\nu \partial_x+u_1))^{-1}$ is the resolvent of the operator $L^*=\nu \partial_x +u_1$
which is adjoint to the ``Lax'' operator $L=-\nu \partial_x +u_1$ for the well known ``Lax pair'' for the 
Burgers equation, i.e.
\begin{equation}
 \left\{
 \begin{array}{l}
  L \psi= (-\nu \partial_x +u_1) \psi\, ,\\
  \psi_t= \nu \psi_{xx}\, .
 \end{array}
 \right. \, .
 \label{Lax-B}
\end{equation}
The formulae (\ref{Grinf}) and (\ref{potinf}) establish the relation between parabolic regularization
by Jordan chain and the method of resolvent developed long time ago for integrable systems \cite{GD1,GD2}.\par
Now let us proceed in an opposite direction. Consider a solution of the Burgers equation (\ref{Burgeq}) and define the infinite set of functions
\begin{equation}
 u_1=\phi\, , \qquad  u_2=\frac{\phi^2}{2}+\nu \phi_x\, , \qquad u_3=\frac{\phi^3}{3}+2 \nu \phi \phi_x +\nu^2 \phi_{xx}\, , \dots
 \label{momenta}
\end{equation}
constructed recursively via
\begin{equation}
 (u_{n+1})_x =   (u_n)_t - \phi (u_{n})_x\,  .
\label{momenta-rr}
 \end{equation}
In terms of these ``momenta'' $u_k$ the Burgers equation becomes the Jordan chain (\ref{inf-Jordan}).
 So the Burgers equation (\ref{Burgeq}) provides us with the class of solutions for the Jordan
 chain with functional dependencies $u_1,u_2,u_3,\dots$.
 The system (\ref{inf-Jordan}) is the form of equation
(\ref{Burgeq}) in terms of countable set of ``momenta''.\par
Momenta $u_k$ defined by (\ref{momenta}) and (\ref{momenta-rr}) have a very simple meaning in terms of the variable $\psi$ converting
the Burgers equation by Cole-Hopf transformation $\phi=\nu \psi_{xx}/\psi$ into the heat equation 
$\psi_t=\nu \psi_{xx}$. Indeed its a simple check 
that the definitions (\ref{momenta}) are equivalent to the following one (see equation (\ref{Lax-B}))
\begin{equation}
 \nu^n \frac{\partial^n \psi/\partial x^n}{\psi}\equiv p_n(u_1,u_2,\dots)\, ,
 \label{colen}
\end{equation}
where $p_n$ are elementary Schur polynomial evaluated at (\ref{momenta}). For $n=1$ it is the Cole-Hopf transformation   while for $n \geq 2$ 
it can be viewed as its extension to the jet space. 
Substituting $\psi=\frac{1}{\nu} \int^x \phi(x',t)\D x'$ into (\ref{colen}), one recovers the formulae 
(\ref{momenta}). \par

The $N$-component Jordan system (\ref{n-Jordan}) is related to the Burgers equation too,  
however with constraints.
Let us take the two-component Jordan system and impose the constraint (\ref{c-JB-n}). The 
first equation of the Jordan system is reduced to the Burgers equation while the the second equation becomes the constraint
\begin{equation}
 \frac{1}{3} {u_1}^3+ 2 \nu {u_1} {u_1}_{x}+\nu^2 {u_1}_{xx}=\mathrm{const} \, .
\end{equation}
Starting with the Burgers equation, introducing $u_1$ and $u_2$ as in (\ref{momenta}) and requiring that $u_1$ and $u_2$ satisfy the 
two-component Jordan system one ends up with the Burgers equation together with the constraint
\begin{equation}
 \frac{1}{3} {\phi}^3+ 2 \nu {\phi}{\phi}_{x}+\nu^2 {\phi}_{xx}=\mathrm{const} \, .
 \label{2-c-B}
\end{equation}
So the two-component Jordan system under the constraint (\ref{c-JB-n}) is equivalent to the Burgers  equation under the constraint 
(\ref{2-c-B}). In a similar way one shows that the $3$-component Jordan system under the constraint (\ref{momenta}) is equivalent to the Burgers 
equation constrained by
\begin{equation}
 \frac{1}{4} {\phi}^4 + 3 \nu \phi^2 \phi_x +\frac{5}{2} \nu^2 \phi_x^2 + 3\nu^2 \phi \phi_{xx}  + \nu^3 {\phi}_{xxx}=\mathrm{const} \, .
 \label{3-c-B}
\end{equation}
In general $N$-component Jordan system under the constraint (\ref{momenta}) is equivalent to the Burgers equation with the constraint 
\begin{equation}
 u_{N+1}(\phi)=\mathrm{const} \, ,
 \label{c-c}
\end{equation}
where $ u_{N+1}(u_1)$ are given in (\ref{momenta-rr}). The constraint (\ref{c-c}) has the simple form in term of the function $\psi$. 
For instance, at $N=2$ it is
\begin{equation}
\frac{\psi_{xxx}}{\psi}+\frac{\psi_x\psi_{xx}}{\psi^2}+\frac{\psi_x^3}{3\psi^3} =\mathrm{const}  \, .
\end{equation}
It is noted that these reductions of the Burgers equation are different from the $N$-phase reductions considered in \cite{MAS}. 
\section{Jordan chain as the universal regularization of the BH equation}
\label{sec-univ}
In our construction of the Jordan chain it arose as the inductive limit when $N \to \infty$ of a $N$ component 
parabolic system with the most degenerate Jordan blocks. Viewed in this way the Jordan chain represents the pure
parabolic regularization of all GCs for the BH equation including, in particular, the Burgers equation.  \par

However if one treats the infinite system (\ref{inf-Jordan}) as a formal chain without insisting on 
parabolicity (like for the MAS chain (\ref{MAS-f})), then the formal Jordan chain offers many other possibilities. Having in mind (\ref{Burgeq}) it is natural to consider a
constraint with a second order derivative. Namely, let us impose the constraint
\begin{equation}
 u_2={u_1}^2 + \frac{1}{4}{u_1}_{xx}\, .
 \label{KdV-con}
\end{equation}
The first equation (\ref{inf-Jordan}) becomes the KdV equation
\begin{equation}
 {u_1}_t=3u_1 {u_1}_x +\frac{1}{4}{u_1}_{xxx}\, ,
 \label{KdV}
\end{equation}
while the other equations define the other fields
\begin{equation}
 u_3= \frac{4}{3} {u_1}^3 +\frac{5}{8} ({u_1}_x)^2 +{u_1}{u_1}_{xx} +\frac{1}{16} {u_1}_{xxxx}\, ,
 \qquad u_4= {u_1}^4 +\dots\, , \qquad \dots \, .
\end{equation}

Moreover, under this constraint the elementary Schur polynomials take the form
$p_k=R^k \cdot 1$ $(k=0,1,2,\dots)$ where $R$ is the KdV recursion operator
\begin{equation}
R=\frac{1}{4} \partial_x^2+u_1 + \partial_x^{-1}(u_1 \partial_x)\, .
\end{equation}
Hence the hierarchy of Jordan chains is reduced to the KdV hierarchy
\begin{equation}
\frac{\partial u_1}{\partial t_k}=\partial_x (R^k \cdot u_1)\, , \qquad k=1,2,3,\dots \, . 
\label{KdVhier}
\end{equation}
In a different way this fact has been first observed in \cite{MAS} for the MAS chain. 

For the reduction (\ref{KdV-con}) the generating function (\ref{Sch-gen}) becomes
\begin{equation}
G(z,u_1)=(1-z R(u_1))^{-1}\cdot 1
\end{equation}
or, formally, 
\begin{equation}
(1-z R(u_1)) \cdot G(z,u_1)= 1\, .
\end{equation}
The KdV constraint  (\ref{KdV-con}) converts the distribution (\ref{N-dens}) into
\begin{equation}
G^{(\infty)}_{\mathrm{KdV}}(u_1,u)=\frac{1}{2 \pi}\int_{-\infty}^{+\infty}\D \lambda
e^{-i\lambda u} (1-i\lambda R(u_1))^{-1}\cdot 1\, ,
\end{equation}
that clearly resembles the resolvent formulae for the KdV equation \cite{GD1,GD2},
while the function (\ref{GATF}) for the Jordan chain hierarchy is reduced to
\begin{equation}
W^{(\infty)}_{\mathrm{KdV}}(u_1;x,t)= \sum_{k=0}^\infty t_k (R^k(u_1) u_1)
+
\frac{1}{2 \pi}\int_{-\infty}^{+\infty}\D \lambda
\tilde{f}(\lambda) (1-i\lambda R(u_1) )^{-1}\cdot 1\, ,
\end{equation}
where$\tilde{f}(\lambda)$ is an arbitrary function, $t_0=x$ and 
$R^k(u_1) \cdot u_1$  with $k=1,2,3,\dots$
are the currents of the conservation laws (\ref{KdVhier}).

On the other hand, starting  with the KdV equation (\ref{KdV}), and introducing the ``momenta'' ($v=u_1$)
\begin{equation}
 u_2= {v}^2 + \frac{1}{4}{v}_{xx} 
 \, , \qquad u_3= \frac{4}{3} {v}^3 +\frac{5}{8} ({v}_x)^2 +{v}{v}_{xx}+\frac{1}{16}v_{xxxx} \, , \qquad \dots \, ,
\end{equation}
{\emph via} the recursion formula 
\begin{equation}
 {u_{n+1}}_x={u_n}_t - v {u_{n}}_x \, , \qquad n=1,2,3,\dots \, ,
\end{equation}
one gets the Jordan chain. So, the KdV equation which provides us a dispersive regularization of the BH equation
is also a particular reduction of the Jordan chain. Many other integrable reductions of the hierarchy of the MAS chain have been found in \cite{MAS,MAS2,Sha}.
However most of them are not gradient catastrophe free. \par
The Jordan chain contains other known regularizations of the BH equation, for instance, Hamiltonian \cite{Dubgen}
and dissipative \cite{DE} ones. For a general dissipative regularization the constraint is \cite{DE}
\begin{equation}
 {u_2}_x= \epsilon \left( b_1(u_1) {u_1}_{xx}+c_1(u_1) ({u_1}_{x})^2 \right) 
 +\epsilon^2 \left( b_2(u_1) {u_1}_{xxx}+c_2(u_1) {u_1}_{x}{u_1}_{xx}+d_2(u_1) ({u_1}_{x})^3 \right) + \dots \, ,
\end{equation}
while for a Hamiltonian regularization \cite{Dubgen}
\begin{equation}
 {u_2}= \epsilon \left( b_1(u_1) {u_1}_{xx}+c_1(u_1) ({u_1}_{x})^2 \right) 
 +\epsilon^2 \left( b_2(u_1) {u_1}_{xxx}+c_2(u_1) {u_1}_{x}{u_1}_{xx}+d_2(u_1) ({u_1}_{x})^3 \right) + \dots \, ,
\end{equation}
where $\epsilon$ is a small parameter and $b_i,c_i,d_i,\dots$ are certain functions \cite{Dubgen,DE}. \par
Other methods of regularization of GCs for the BH equation are also related to the Jordan chain. One of the
approaches to regularization of GC is to incorporate the derivatives and pass from the equation for critical 
points of functions to Euler-Lagrange equations for functionals (see e.g. \cite{KMAM,KO}). In order to regularize the 
$k$-th order GC for the BH equation one passes from the function  $W^*_k$ (\ref{potbalBH-star}) to the ``Lagrangian''
\begin{equation}
 {\cal L}_k= \frac{a}{2} ({\upsilon}_{{{y}}})^2 + W^*_k\, ,
 \label{Lagk}
\end{equation}
where $a$ is a constant. Euler Lagrange equations for ${\cal L}_k$  is
\begin{equation}
 a u_{{{y}}{{y}}} - {{y}} - {\tau} {\upsilon} -(k+2) A_{k+2}{{\upsilon}}^{k+1}=0\, .
 \label{ELagk}
\end{equation}
Solutions of this equation obviously do not exhibit GC. 
For $k=1$ the equation (\ref{ELagk}) is a sort of deformed Painlev\'e I equation. At $k=2$ this equation, i.e.
\begin{equation}
 a u_{{{y}}{{y}}} - {{y}} - {\tau} {\upsilon} -4 A_{4}{{\upsilon}}^{3}=0\, ,
 \label{ELagk-2}
\end{equation}
is the equation obtained in \cite{Sul} within the singularity analysis of the diffusion type equation
and then studied in \cite{IS}. For $k>2$ equations (\ref{ELagk}) have been obtained in \cite{KMAM}
as a particular case of a systems of PDEs. \par

Equations (\ref{ELagk}) represent themselves the regularization of the GC for the BH equation of orders $k=1,2,3,\dots$.
On the other hand, all of them provide us with particular classes of solutions (with different $k$) of a single PDE.
Similarly to the hodograph equation (\ref{solhodoBH}) one obtains this equation differentiating (\ref{ELagk}) w.r.t. ${{y}}$ and 
${\tau}$ and eliminating ${\tau}$ from the obtained system. In the case of equation (\ref{ELagk}) one gets the equation
\begin{equation}
 {\upsilon}_{{\tau}}-{\upsilon}{\upsilon}_{{{y}}} +a ({\upsilon}_{{{y}}{{y}}{\tau}}{\upsilon}_{{{y}}}-{\upsilon}_{{{y}}{{y}}{{y}}} {\upsilon}_{{\tau}})=0\, .
 \label{nonevreg}
\end{equation}
or, equivalently
\begin{equation}
(1-a \partial_{{{y}} }({\upsilon}_{{{y}}{{y}}}-{\upsilon}_{{{y}}} \partial_{{{y}}})) {\upsilon}_{{\tau}}={\upsilon}{\upsilon}_{{{y}}}\, .
 \label{nonevreg-I}
\end{equation}
For small $a$ equation (\ref{nonevreg-I}), at first order in $a$, looks like
\begin{equation}
 {\upsilon}_{{\tau}}={\upsilon}{\upsilon}_{{{y}}}-3a ({\upsilon}_{{{y}}})^2 {\upsilon}_{{{y}}{{y}}} \, .
 \label{reg-small}
\end{equation}
Equation (\ref{reg-small}) is a very special case of general Hamiltonian deformations of the BH equation
studied in \cite{Dubgen}. Complete equation (\ref{nonevreg-I}) is the Jordan chain under the constraint 
\begin{equation}
 u_1={\upsilon_1}\, , \qquad  {u_2}_x=(a {{u_1}_{xxx}}-a {u_1}_x \partial_x^2){u_1}_t\, .
\end{equation}
One has different regularization of GCs for the BH equation passing from  $W^*_k$ (\ref{potbalBH-star})
to the Lagrangian 
\begin{equation}
 \overline{{\cal L}}_k= \frac{a}{2} {\upsilon}_{{{y}}}{\upsilon}_{{\tau}} + W^*_k\, .
 \label{Lagk-nev}
\end{equation}
Euler Lagrange equations for $\overline{{\cal L}}_k$ gives
\begin{equation}
 a u_{{{y}}{\tau}} - {{y}} - {\tau} {\upsilon} -(k+2) A_{k+2}{{\upsilon}}^{k+1}=0\, .
 \label{ELagk-nev}
\end{equation}
All the equations (\ref{ELagk-nev}) with ($k=1,2,3,\dots$) provide subclass of solutions for the equation
\begin{equation}
 {\upsilon}_{{\tau}}-{\upsilon}{\upsilon}_{{{y}}} +a ({\upsilon}_{{{y}}{\tau}{\tau}}{\upsilon}_{{{y}}}-{\upsilon}_{{{y}}{{y}}{\tau}} {\upsilon}_{{\tau}})=0\, ,
 \label{nonevreg-ne}
\end{equation}
which is again a suitable reduction of the Jordan chain. \par
Similar type of regularization can be performed also for $N$-components Jordan systems. One deforms
the functions ${W^{(N)}}^*$ (\ref{pot-n-J-k-star}) to the Lagrangian
\begin{equation}
 {\cal L}^{(N)}=\frac{a}{2} ({{\upsilon_1}}_{{{y}}})^2+{W^{(N)}}^*\, ,
\end{equation}
taking into account that the derivative  ${{\upsilon_1}}_{{{y}}}$ is the most singular one.
One gets the following equations 
\begin{equation}
 \left\{
\begin{array}{l}
 a {{\upsilon_1}}_{{{y}}{{y}}} - {{y}} - {\tau} {\upsilon}_1 + A_{N+k+1}P_{n+k}({\upsilon})=0\, , \\
 {\tau} - A_{N+k+1}P_{n+k-1}({\upsilon})=0\, , \\
 P_{n+k-l+1}({\upsilon})=0\, , \qquad l=3,4,\dots,N\, .\\
\end{array}
 \right.
\end{equation}
These equations represent a completely degenerate parabolic regularization of the $k$-th order GC
for the $N$-component Jordan system. Of course, there are number of other regularizations
according to different choices of differential part for the Lagrangian ${\cal L}^{(N)}$.\par
So the Jordan chain can be viewed as the universal regularization of the BH equation. Specific
regularizations correspond to its particular reductions. Some of these reductions lead to integrable PDEs
like the Burgers and KdV equations for which the Jordan chain hierarchy is reduced to the infinite 
family of commuting flows. Characterization of such reductions and associated physical phenomena is an
interesting open problem.
\subsubsection*{Acknowledgments} One of the authors (BK) thanks Y. Kodama and M. Pavlov  for fruitful discussions. The authors thank the anonymous referee for useful suggestions. 
This work was carried out under the auspices of the GNFM Section of INdAM.

\appendix
\section{First step of the regularization procedure}
\label{app-expl}
Here we present some explicit formulae relating the BH equation and the two-component Jordan system.
Let us consider the BH equation with the initial data
\begin{equation}
u(x,0)=u_0(x)\, .
 \label{inidaBH}
\end{equation}
Solution of this Cauchy problem is given implicitly by the hodograph equation
\begin{equation}
W_u \equiv x+u t +\wW_u(u)=0
 \label{hodoBH}
\end{equation}
where $-\wW_u(u)$ in the inverse function of $u_0(x)$. Differentiating (\ref{hodoBH}) w.r.t. $x$ and $t$ one gets
\begin{equation}
 1+W_{uu} u_x=0\, , \qquad u + W_{uu} u_t=0\, .
\label{hodoBHxt}
\end{equation}
At the regular sector ($W_{uu} \neq 0$) one has the formulae (\ref{derGC1}). First order GC appears when 
\begin{equation}
 W_{uu}=t + \wW_{uu}=0\, .
 \label{GC1BH}
\end{equation}
Equations (\ref{hodoBH}) and (\ref{GC1BH}) defined the submanifold of codimension one in the space $x,t$ and 
parameters $t_2,t_3,\dots$ (see \cite{Arn1}).\par
In order to regularize the first GC of the BH equation we enlarge the space of variables $u \to (u_1,u_2)$ and pass to the system
(\ref{2genBHeq}). This system describes the dynamics of the critical points for the functions (\ref{pot2}), i.e. 
\begin{equation}
  W^{(2)}(x,t,u_1,u_2)=x u_1+t\left( \frac{u_1^2}{2}+u_2\right)+\wW^{(2)}(u_1,u_2)
 \label{pot2-app}
\end{equation}
which obey the equation (\ref{heat})
\begin{equation}
 W^{(2)}_{u_2}=W^{(2)}_{u_1u_1}.
\end{equation}
The equations for the critical points are
\begin{equation}
 x+t u_1+\wW^{(2)}_{u_1}=0\, , \qquad t +\wW^{(2)}_{u_2}=0\, .
 \label{hodo2}
\end{equation}
Differentiating these equations w.r.t. $x$ and $t$ one gets
\begin{equation}
 \begin{split}
& 1+ W^{(2)}_{u_1 u_1} {u_1}_x+ W^{(2)}_{u_1 u_2} {u_2}_x=0\, , \qquad W^{(2)}_{u_2 u_1} {u_1}_x+W^{(2)}_{u_2 u_2} {u_2}_x=0\, , \\
& u_1 + W^{(2)}_{u_1 u_1} {u_1}_t+W^{(2)}_{u_1 u_2} {u_2}_t=0\, , \qquad 1 +W^{(2)}_{u_2 u_1} {u_1}_t+W^{(2)}_{u_2 u_2} {u_2}_t=0\, . \\
 \end{split}
\label{pot2crit-ugly}
\end{equation}
Taking into account the second equation (\ref{hodo2}), one finally obtains
\begin{equation}
 \begin{split}
& 1+W^{(2)}_{u_1 u_1 u_1} {u_2}_x=0\, , \qquad W^{(2)}_{u_1 u_1 u_1} {u_1}_x+W^{(2)}_{u_1 u_1 u_1 u_1} {u_2}_x=0\, , \\
& u_1 +W^{(2)}_{u_1 u_1 u_1} {u_2}_t=0\, , \qquad 1 +W^{(2)}_{u_1 u_1 u_1} {u_1}_t+W^{(2)}_{u_1 u_1 u_1 u_1} {u_2}_t=0\, . \\
 \end{split}
\label{pot2crit}
\end{equation}
At he regular sector when $W^{(2)}_{u_1 u_1 u_1} \neq 0$ one has
\begin{equation}
  \begin{split}
& {u_2}_x=- \frac{1}{W^{(2)}_{u_1 u_1 u_1} }\, , \qquad  {u_1}_x =-\frac{W^{(2)}_{u_1 u_1 u_1 u_1}}{(W^{(2)}_{u_1 u_1 u_1})^2}\, , \\
& {u_2}_t=-\frac{u_1}{W^{(2)}_{u_1 u_1 u_1}}\, , 
\qquad  {u_1}_t=-\frac{u_1 W^{(2)}_{u_1 u_1 u_1 u_1}}{(W^{(2)}_{u_1 u_1 u_1})^2}-\frac{1}{ W^{(2)}_{u_1 u_1 u_1}}\, . \\
 \end{split}
\label{der2crit}
\end{equation}
Comparing equations (\ref{hodoBHxt}) and (\ref{pot2crit}), (\ref{der2crit}), one notes that in the transition from the
one-component case to two-component one, even if the condition $W_{uu}=0$ (defining first order GC for the BH equation)
is substituted by the similar condition ${W^{(2)}_{u_1 u_1} }=0$, the later one does not lead to the blow up of the derivatives of $u_1$
and $u_2$. In contrast to the equations (\ref{hodoBHxt}) the second equation (\ref{hodo2}) does not define the singular sector, but
together with the first equation serves to calculate the two functions $u_1(x,t)$ and $u_2(x,t)$.  First GC for the system (\ref{2genBHeq})
happens when ${W^{(2)}_{u_1 u_1 u_1} }=0$ and the derivative blow up according to the  formulae (\ref{der2crit}). \par

Transition from the BH equation to the two-component Jordan system can be performed averaging with the probability distribution
(\ref{heatfund}) namely
 \begin{equation}
\begin{split}
 & u_1=\langle u\rangle_2 = \frac{1}{\sqrt{4 \pi u_2}} \int_{-\infty}^{+\infty}  \D  u \, u \exp \left({-\frac{(u_1-u)^2}{4u_2}}\right) \,  ,\\
 & u_2=\frac{1}{2} (\langle u ^2\rangle_2- \langle u\rangle_2^2) =
 \frac{1}{2\sqrt{4 \pi u_2}} \int_{-\infty}^{+\infty}  \D  u \,  {u^2} \exp\left({-\frac{(u_1-u)^2}{4u_2}}\right)-
 \frac{1}{2}u_1^2
 \, , \\
 & W^{(2)}(u_1,u_2)= \langle W(u)\rangle_2
 =\frac{1}{\sqrt{4 \pi u_2}} \int_{-\infty}^{+\infty}  \D  u \, W(u) \exp \left({-\frac{(u_1-u)^2}{4u_2}}\right)  \,  ,\\
 & \frac{\partial^k W^{(2)}(u_1,u_2)}{\partial u_1^k}= 
 \langle \frac{\partial^k W(u)}{\partial u^k}\rangle_2=\frac{1}{\sqrt{4 \pi u_2}} \int_{-\infty}^{+\infty}  \D  u \, 
 \frac{\partial^k W(u)}{\partial u^k} \exp\left({-\frac{(u_1-u)^2}{4u_2}}\right)  \,  , \quad k=1,2,3,\dots\, .
\end{split}
\label{aver2}
 \end{equation}
Averaging of equations (\ref{hodoBH}) and (\ref{GC1BH}) for the BH equation gives equations (\ref{hodo2}) defining critical points
of $W^{(2)}(u_1,u_2)$. Thus this averaging regularize the first order GC for the BH equation. \par

Then the last equation of (\ref{aver2}) with $k=3$, i.e.,
\begin{equation}
  \frac{\partial^3 W^{(2)}(u_1,u_2)}{\partial u_1^3}= 
 \langle \frac{\partial^3 W(u)}{\partial u^3}\rangle_2=\frac{1}{\sqrt{4 \pi u_2}} \int_{-\infty}^{+\infty}  \D  u \, 
 \frac{\partial^3 W(u)}{\partial u^3} \exp\left({-\frac{(u_1-u)^2}{4u_2}}\right)  
 \label{potBH-J}
\end{equation}
implies that $W^{(2)}_{u_1 u_1 u_1}$ vanishes at some point $({u_1}_0,{v_1}_0)$ only if  $W_{uuu}$ takes positive and negative values.
Assuming its smoothness, one concludes that there exists a point $u_0$ at which $W_{uuu}(u_0)=0$ and $W_{uuuu}(u_0)\neq0$.  For the BH equation
such condition is realizable at the second stratum $S_2$ which exhibits the second order GC. So the first order GC for the system (\ref{2genBHeq}) 
is related to the second order GC of the BH equation.
Analogously, the last formula (\ref{aver2}) with $n>3$ establishes the relation between $n$-th order GC for the BH equation
and the GC of the order $n-1$ for the two-component Jordan system. \par
The process of averaging described above admits a simple physical interpretation. Let us consider the BH equation as the equation which 
describes the collisionless motion of a cloud of particles of dust with mass $m=1$
(see e.g. \cite{Zel,SZ}). In addition to the usual picture we assume that the cloud is an ideal Boltzmann gas at temperature $T$. At $T=0$
one has the flow of particles with the velocities $u(x,t)$ and GC of first order. At $T>0$  the particles have different values of
velocities with the probability given by the Maxwell distribution
\begin{equation}
 \rho(u,u_1,T)=\frac{1}{\sqrt{2 \pi T}} \exp \left( -\frac{(u-u_1)^2}{2 T}\right)\, .
 \label{Max}
\end{equation}
It is the Gau{\ss} distribution (\ref{heatfund}) with $u_2=T/2$. The first two formulae (\ref{aver2})
are the classical statistical physics formulae, namely $u_1=\langle u \rangle$ is the velocity of 
macroscopic motion of the cloud and $E=\langle \frac{u^2}{2} \rangle=\frac{1}{2}u_1^2 +\frac{T}{2}$ 
is the mean value of the single particle energy. 
We assume also that both macroscopic quantities $u_1$ and $T$ depend on $x$ and $t$ in a way that the 
function $W^{(2)}= x u_1 + t E +\wW(u_1,T)$ has the extremum. 
So one has the system (\ref{2genBHeq}) with $u_2=T/2$. Thus, the first order GC for the dust 
particles motion is regularized by the thermal effect of nonzero and variable temperature $T$.


\end{document}